\documentclass{aastex631}

\begin{document}

\title{Scalable, Advanced Machine Learning-based Approaches for Stellar Flare Identification: Application to TESS short-cadence Data and Analysis of a New Flare Catalogue}

\author[0000-0001-5989-7594]{Chia--Lung Lin}
\correspondingauthor{Chia--Lung Lin}
\email{m1059006@gm.astro.ncu.edu.tw, chialunglin@arizona.edu}
\affiliation{Graduate Institute of Astronomy, National Central University, Taoyuan 32001, Taiwan}
\affiliation{Steward Observatory, The University of Arizona, Tucson, AZ 85721, USA}

\author[0000-0003-3714-5855]{D\'aniel Apai}
\affiliation{Steward Observatory, The University of Arizona, Tucson, AZ 85721, USA}
\affiliation{Lunar and Planetary Laboratory, The University of Arizona, Tucson, AZ 85721, USA}

\author[0000-0002-2132-5264]{Mark S. Giampapa}
\affiliation{Steward Observatory, The University of Arizona, Tucson, AZ 85721, USA}
\affiliation{Lunar and Planetary Laboratory, The University of Arizona, Tucson, AZ 85721, USA}
\affiliation{National Solar Observatory, 950 N. Cherry Avenue, Tucson, AZ 85719, USA}

\author[0000-0002-3140-5014]{Wing--Huen Ip}
\affiliation{Graduate Institute of Astronomy, National Central University, Taoyuan 32001, Taiwan}
\affiliation{Graduate Institute of Space Science, National Central University, Taoyuan 32001, Taiwan}



\received{2024 June 12}
\revised{2024 August 15}
\accepted{2024 August 30}
\submitjournal{The Astronomical Journal}



\begin{abstract}

We apply multi-algorithm machine learning models to TESS 2-minute survey data from Sectors 1-72 to identify stellar flares. Models trained with Deep Neural Network, Random Forest, and XGBoost algorithms, respectively, utilized four flare light curve characteristics as input features. 
Model performance is evaluated using accuracy, precision, recall, and F1-score metrics, all exceeding 94\%. 
Validation against previously reported TESS M dwarf flare identifications showed that our models successfully recovered over 92\% of the flares while detecting $\sim2,000$ more small events, thus extending the detection sensitivity of previous work.
After processing 1.3 million light curves, our models identified nearly 18,000 flare stars and 250,000 flares. 
We present an extensive catalog documenting both flare and stellar properties. 
We find strong correlations in total flare energy and flare amplitude with color, in agreement with previous studies. Flare frequency distributions are analyzed, refining power-law slopes for flare behavior with the frequency uncertainties due to the detection incompleteness of low-amplitude events. We determine rotation periods for $\sim120,000$ stars thus yielding the relationship between rotation period and flare activity.  We find that the transition in rotation period between the saturated and unsaturated regimes in flare energy coincides with the same transition in rotation period separating the saturated and unsaturated levels in coronal X-ray emission.
We find that X-ray emission increases more rapidly with flare luminosity in earlier-type and unsaturated stars, indicating more efficient coronal heating in these objects. 
Additionally, we detect flares in white dwarfs and hot subdwarfs that are likely arising from unresolved low-mass companions.

\end{abstract}


\section{Introduction} \label{sec:intro}




Stellar flares are intense, sudden outbursts of energy on the surfaces of stars, caused by the reconnection of magnetic fields \citep[e.g.,][]{1963ApJS....8..177P}. 
This magnetic reconnection process rapidly heats the stellar atmosphere and accelerates plasma stored in magnetic loops, releasing energy across a broad spectrum, from X-rays to radio wavelengths in addition to energetic particles \citep[][]{2005ApJ...621..398O, 2015A&A...581A..28P, 2019ApJ...871..167K, 2023MNRAS.519.3564J}.
The primary Kepler mission and the later K2 mission have greatly advanced the study of flares, discovering many superflares on stars of various spectral types 
\citep{2012Natur.485..478M, 2012MNRAS.423.3420B, 2013ApJS..209....5S, 2014ApJ...797..121H, 2015ApJ...798...92W, 2018ApJ...867...78C, 2019ApJS..241...29Y, 2019ApJ...873...97L}.
After the end of the Kepler/K2 mission, the Transiting Exoplanet
Survey Satellite \citep[TESS,][]{2015JATIS...1a4003R} extended this work with its short-cadence observations, and several studies have carried out the detection of flares over a wide range of stars from A-type to ultra-cool dwarfs \citep{2020AJ....159...60G, 2020ApJ...905..107M, 2023A+A...669A..15Y, 2024MNRAS.527.8290P}.

These missions have revealed that flares occur on nearly all main-sequence stars, particularly those with outer convective envelopes. 
Observational evidence indicates that low-mass or late-type main-sequence stars (i.e., M, K, and G-type dwarfs) exhibit higher flare frequencies than more massive stars \citep[e.g.,][]{2019ApJ...873...97L, 2023A+A...669A..15Y}.
In addition, flare frequency and energy are correlated with the stellar rotation period, which is also an indicator of stellar age, a relationship known as gyrochronology \citep{1972ApJ...171..565S, 2007ApJ...669.1167B}.
Fast rotating/younger stars tend to be more magnetically active than older stars that generally spin more slowly \citep[e.g.,][]{2016ApJ...829...23D}.
Moreover, superflares with energies more than \(10^{33}\)~ergs can be generated by low-mass stars \citep[e.g.,][]{2000ApJ...529.1026S, 2021AJ....162...11L}.  
The powerful energy and high frequency of the flares of the low-mass stars could significantly impact habitability by inducing biologically adverse chemical reactions in the atmospheres of surrounding exoplanets \citep{2010AsBio..10..751S, 2021NatAs...5..298C}.
Recently, researchers have recognized that flare activity in low-mass stars can interfere with exoplanet detectability and the interpretation of transmission spectra. 
These flares could contaminate observations used for detecting exoplanets and determining their properties, whether in time-domain photometric data for planet detection, spectroscopic data for mass and radial velocity measurements, or transmission spectra for analyzing planetary atmospheres \citep{2018ApJ...853..122R, 2018AJ....156..178Z, 2023AJ....165..149D, 2023ApJ...959...64H}.

In recent years, several machine learning and Bayesian statistic-based models have been developed for detecting flares in Kepler, K2, and TESS light curves, each demonstrating varying degrees of efficiency. 
\cite{2014MNRAS.445.2268P} introduced \texttt{Bayesflares}, a Bayesian-odds-ratio-based algorithm, for detecting stellar flares in Kepler data. This model, featuring a rapid rise and exponential decay combined with a polynomial background to address light curve variations, successfully identified 687 flaring stars with a total of 1873 flares in Kepler Quarter 1 data. However, the software was developed in \texttt{Python2}, which is now outdated and no longer supported.
\cite{2022AJ....163..147G} developed a modified version of \texttt{Bayesflares} and successfully identified several flares in AU~Mic by using 20-second and 2-minute cadence TESS data.
Despite its success, this modified version has not been made public.

By using the RANdom SAmple Consensus algorithm to robustly model light curves with flares, \cite{2018A&A...616A.163V} presented the first machine learning based code, \texttt{Flatwrm}, to search for flares in Kepler data, but it was programmed in the now outdated \texttt{Python2}. 
They later introduced an advanced version, \texttt{Flatwrm2} \citep{2021A&A...652A.107V}, which employs recurrent deep neural networks in \texttt{Python3}, though its application is limited to Kepler/K2 and TESS data with the specific observation cadences.

\cite{2020AJ....160..219F} created a convolutional neural network (CNN) flare searching model, \texttt{stella}, specifically for the TESS short-cadence data only. They used this network to analyze 3200 young stars effectively, assessing flare rates based on ages and spectral types. 
By also using a CNN algorithm, \cite{2022ApJ...935...90T} developed image-based classification models to identify superflares in solar-type stars from TESS Target-Pixel-file images, unlike the previous models that focused solely on light curve data.
However, their method is primarily effective for high-amplitude superflares and less so for the detection of low-amplitude and short-duration events. 
The complexity of the image classification model demands substantial computational resources for training and executing classification tasks.

To sum up, machine learning has been instrumental in developing various models for flare identification. 
However, due to issues of obsolescence, efficiency, cadence limitation, and considering that all existing models rely heavily on a single algorithm, we introduce in this work a new machine-learning approach that incorporates multiple algorithms: Deep Neural Network (DNN), Random Forest (RF), and XGBoost. 
This innovative method aims to enhance effectiveness and sensitivity in identifying flares from light curves observed at various cadences, while maintaining a low false positive rate.
We train these models using the TESS 2-minute short cadence data and apply them to the entire TESS short cadence dataset available as of January 2024, which consists of approximately 1.3 million light curves. 
By doing so we significantly increase the detection of flare events thus enabling us to extend previous work on flare properties as a function of spectral type to lower flare amplitudes.

The structure of this paper is organized as follows: Section~\ref{sec:dataset pre-processing and sample collection} provides an overview of the TESS observations and details the data preprocessing steps, sample collection, and the definition of characteristics for the flare light curve profiles used in our pre-classified sample set. In Section~\ref{sec:models_traning}, we introduce the foundational concepts of the Deep Neural Networks, Random Forest, and XGBoost algorithms, and describe the training process of our models. Section~\ref{sec:model_performace_assessment} evaluates the performance of our flare identification approach through a flare recovery task, comparing our results with existing flare detections in the literature. Section~\ref{sec:deployment_on_tess_data} illustrates the application of our models to the TESS short cadence dataset along with the results of our flare detection efforts. 
In Section~\ref{sec:discussion}, we explore the properties of flares across different spectral types of stars. Finally, Section~\ref{sec:conclusion} concludes our study and discusses potential directions for our future research.


\section{Dataset Pre-processing and sample collection from TESS short cadence observation}
\label{sec:dataset pre-processing and sample collection}
TESS (the Transiting Exoplanet Survey Satellite), a NASA mission led by MIT (PI: Dr. George R. Ricker) and launched on April 18, 2018, aims to discover nearby transiting exoplanets \citep{2015JATIS...1a4003R}. 
It utilizes two-minute short-cadence time-series data with high photometric precision and a wavelength coverage of approximately 600--1000~nm, making it ideally suitable for stellar flare studies.
The four cameras, with the pixel scale of $21''\times 21''$, of TESS provide a combined field of view of $24^{o} \times 96^{o}$, effectively covering the sky in two hemispheres divided into 13 sectors each. 
Each sector represents a block within the TESS field of view, with each block being observed for approximately 27 days. 
TESS sequentially observes these blocks, moving from one to the next, until it covers all sectors.
By January, 2024, TESS have observed 72 sectors covering about 85\% of the sky and accumulating approximately 1.3 million short-cadence light curves from over 490,000 sources.

In this section, we illustrate the approaches to pre-processing the TESS short cadence light curves and display how we collect true and false flare samples to train the machine learning models.
To carry out our study, we utilized the Python package, "\texttt{Lightkurve}" \citep{2018ascl.soft12013L},  which specializes in the Kepler/K2 and TESS light curve access and analysis. 
To efficiently obtain true flare samples, our primary focus is on the well-documented flaring low--mass stars, such as Proxima Centauri, Wolf~359, GJ~3147, etc., reported in the literature \citep{1999A&AS..139..555G, 2019ApJ...884..160V, 2020AJ....159...60G, 2021AJ....162...11L}.
A total of 15 stars, which serve as sample sources along with relevant information on their TESS data sectors, are listed in Table~\ref{tab:16_stars_sample_sources}.

The TESS dataset includes two types of photometric flux measurements: the Simple Aperture Photometric (SAP) flux and the Pre-search Data Conditioning Simple Aperture Photometric (PDCSAP) flux. The SAP flux is computed by aggregating the calibrated pixel counts within an aperture defined by the TESS Science Processing Operations Center (SPOC) pipeline \citep{2016SPIE.9913E..3EJ}. However, SAP flux measurements are susceptible to systematic errors, including long-term trends. These errors are mitigated in the PDCSAP flux through the application of Co-trending Basis Vectors, rendering the PDCSAP data typically more refined in comparison to SAP flux. For the purposes of this study, we have utilized the PDCSAP flux, specifically selecting data points with a 'good quality' flag (flag bits QUALITY$=$0) for our analysis.

\subsection{Flare Candidate Identification}
\label{subsec:flare candidates identification}
For each light curve, we removed the long-term brightness variation, such as spot modulation, by using the \texttt{flatten} function feature of the \texttt{Lightkurve} package with customized parameters. 
{This process was conducted in two iterations to circumvent outliers, such as flares and transits, by using the mask parameter in the \texttt{flatten} function, which excludes unwanted data points from the second-order polynomial spline function used for light curve detrending.}
In the initial iteration, we executed the \texttt{flatten} function without giving the mask parameter. 
Following this, we calculated the standard deviation ($\sigma_{fr}$) of the first-round flattened light curve and defined the mask parameter for the \( i \)-th data point as "True" if the flux of that data point was greater than three times $\sigma_{fr}$.
Subsequently, we reapplied the \texttt{flatten} function, this time with the designated mask parameter, to obtain a second-round flattened light curve. This refined light curve (hereafter $LC_{1}$) would be utilized for further analysis.

The flux of the $LC_{1}$ has been normalized by using the following equation
\begin{equation}
  \frac{\Delta F(t)}{\widetilde{F}} = \frac{F(t)-\widetilde{F}}{\widetilde{F}}, 
  \label{eq:eq_amplitude}
\end{equation}
where $F(t)$ is the flux as a function of time, and $\widetilde{F}$ is the median of a star’s quiescent flux.  
We employed the following criteria, which is a modified version of criteria from \cite{2015ApJ...814...35C} and \cite{2019ApJ...873...97L} to select flare candidates in \( LC_{1} \)
\begin{enumerate}
  \item \( F_{\text{LC}_{1i}} > 0 \), where \( F_{\text{LC}_{1i}} \) is the normalized flux for the \( i \)-th data point.
  \item \( \frac{F_{\text{LC}_{1i}}}{{\sigma}(F_{\text{LC}_{1}})} \geq 1 \), where \( {\sigma}(F_{\text{LC}_{1}}) \) is the standard deviation of the normalized flux. This standard deviation is computed while excluding mostly outliers by applying the mask parameters defined for the second light curve flattening iteration.
  Empirically, the typical value for this criterion is either 2 or 3. However, we have set it to 1 in our study to broaden the diversity of our training dataset. 
  \item \( \text{Cons} \geq 3 \), indicating that the count of consecutive data points meeting the above criteria must be at least three. 
  {Flares are rapid events, with durations as short as 1 minute \citep[e.g.,][]{2021ApJS..253...35T} or even shorter. 
  Spikes due to cosmic ray hits or noise fluctuations usually appear in only one or two data points. 
  Therefore, it is impossible to detect flares with very short duration using 2-minute cadence TESS data. 
  We set the threshold to three consecutive points to maximize flare detection while minimizing misidentification due to cosmic ray hits or noise fluctuations.}
\end{enumerate}
In this way, we gathered 18,075 flare candidates from the short-cadence TESS data of 15 stars. 
The start time of each candidate is defined as the recorded time of the first data point ahead of the first data point with the flux that exceeds the 1~$\times \sigma(F_{\text{LC}_{1}})$ criterion. The end time is marked by the time recorded for the second data point following the final point with the flux surpassing 1~$\times \sigma(F_{\text{LC}_{1}})$.

We further verified the flare candidates' authenticity by visually examining both their one-dimensional light curve profiles and the corresponding Target Pixel File (TPF) images. 
{We inspected the light curve shape and characteristics of the flares. 
True flares typically exhibit a rapid rise and a slower decay, whereas false positives, such as cosmic ray hits, may show sudden, sharp spikes with little to no decay phase.}
TPF images were used to extract the light curve profiles from individual pixels within the aperture as defined by the SPOC pipeline, allowing us to determine whether cosmic ray hits and/or background sources may have influenced the flare candidate signals.
{Our process for "True/False" flares determination using TPF data follows the methodology described in \cite{2015ApJ...798...92W}, particularly Figure 3 and Section 2 on pages 4 to 5 in their paper, and in \cite{2023AJ....166...82L}, particularly Figure 2 of their paper. 
These steps allow us to ensure that the flare signal was isolated and originated from the target star rather than neighboring stars or cosmic ray events.}

As the result, these 15 stars are not affected by the bright background and nearby sources too significantly.
Also, we found that the remaining data with a good quality flag already effectively excluded cosmic ray interference, as evidenced by the near absence of such signals during the TPF image inspections. 
{Moreover, to assess the contamination of nearby sources in the TESS data more effectively, we utilized the contamination ratio ($R_{\text{cont}}$) from the Revised TESS Input Calatog as outlined by \cite{2019AJ....158..138S}.
For stars without a reported $R_{\text{cont}}$, we estimated this parameter using the open-source code \texttt{TIC\_CONTAM.PY} by \cite{2021arXiv210804778P}, which adopts Stassun et al.'s methodology. 
This process involves identifying all point sources within 10 TESS pixels of the target with TESS magnitudes between 17 and 19. 
Their fluxes are computed based on pre-launch PSF measurements at the field center, and the size and shape of the target’s aperture are tailored to its TESS magnitude. 
$R_{\text{cont}}$ is then determined as the ratio of the flux from these objects within the aperture to the flux from the target itself.
We highlight that TESS data for stars with $R_{\text{cont}} > 0.1$ are likely contaminated, and we consider the flare amplitudes and energy reported in these data as lower limits. 
Most of these 15 stars have $R_{\text{cont}} < 0.1$, except for TIC 388857263, TIC 220433363, TIC 231914259, and TIC 441398770. 
However, these stars are not significantly affected (e.g., the maximum $R_{\text{cont}}$ in these stars is 0.8 for TIC 388857263, Proxima Centauri), and they are all known eruptive variables. Including these stars may enhance the diversity of the data used to train the models, allowing for better identification of flares in potentially contaminated sources.
Therefore, employing the TPF images to further verify the flares identified by our trained machine-learning algorithm models won't be necessary. 
Additionally, instead of removing these samples from our flare catalog, we will retain the flares detected in stars with $R_{\text{cont}} > 0.1$ and add a flag to indicate potential contamination.}

In summary, we confirmed 4,589 of 18,075 candidates as True flare events, with the remainder classified as non-flare (False) events. 
However, this dataset is too imbalanced for pre-classification in machine learning with a ratio between "True" and "False" of about 1:3. 
We outline in Section~\ref{subsec:oversampling} the oversampling technique employed to produce a balanced pre-classified dataset. 
This will be discussed after explaining the definitions of the four key classification characteristics of the flare profile in the following section.

\subsection{Definition of Flare Profile Characteristics}
\label{subsec:flare_classification_feature}
{Typically, machine learning models rely on quantifiable features defined by people to learn from the data. In our case, features derived from the flare light curve are essential for the model to recognize and classify flares effectively. Therefore, it is crucial to define an appropriate set of characteristics, or features, that reflect the structure of the flare light curve profile for developing machine-learning models for flare identification. These features directly influence the algorithm's ability to learn patterns and make accurate classifications.}

We first gathered a couple of physical properties that can be derived from the flare light curve profile. They are the time duration ($\delta t$), the impulsive phase time length ($\delta t_{1}$), the decay phase time length ($\delta t_{2}$), the equivalent duration (ED), and peak amplitude ($A_{f}$) of the flare. The equivalent duration of the flare can be estimated by using the following equation \citep{1972Ap&SS..19...75G}:
\begin{equation}
  ED = \int \frac{\Delta F(t)}{\widetilde F} dt . 
  \label{eq:ED}
\end{equation}
It is the amount of time required for the star’s quiescence to emit the same amount of energy as the cumulative energy of the given flare.
Next, we defined four characteristics on the flare light curve profile based on these properties: 
\begin{enumerate}
    \item The duration of the flare ($\delta t$) in days.
    \item The impulsive-decay time ratio ($\delta t_1 / \delta t_2$).
    \item The upper-to-lower equivalent duration ratio ($ED_1 / ED_2$).
    \item The peak amplitude signal-to-noise ratio ($A_f / \sigma(F_{LC1})$).
\end{enumerate}

\begin{figure}
    \centering
    \epsscale{0.7}
    \plotone{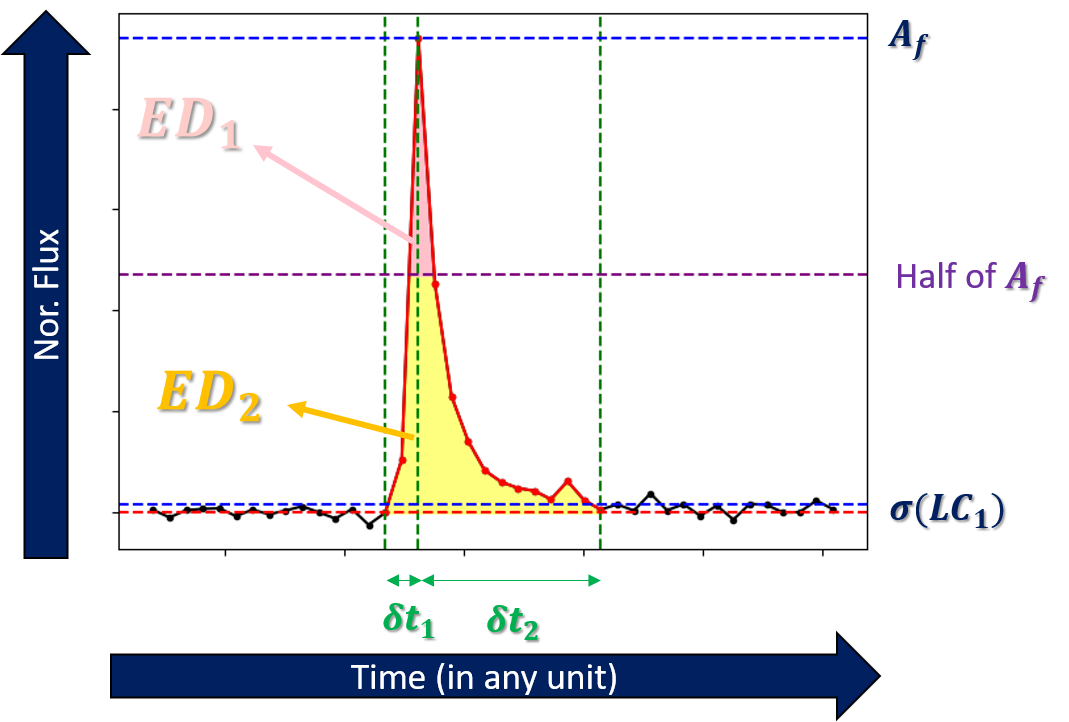}
    \caption{The illustration of the four classification characteristics in the flare light curve profile. The blue horizontal dashed lines from the top to the bottom mark the peak amplitude ($A_{f}$) and the $\sigma(F_{\text{LC}_{1}})$. 
    The red horizontal dashed line represents the median flux of the light curve.
    The purple horizontal dashed line represents half of the peak amplitude. The highlighted areas separated by the purple line represent the regions under the curve used to calculate \( ED_{1} \) and \( ED_{2} \). 
    The green vertical dashed lines from left to right mark the time positions of the flare start, peak, and end, with the impulsive rise and decay time intervals indicated as \( \delta t_{1} \) and \( \delta t_{2} \), respectively. 
    }
    \label{fig:classification_char}
\end{figure}

Figure~\ref{fig:classification_char} visualizes the illustration of these characteristics.  
The upper and lower equivalent duration, denoted as \( ED_{1} \) and \( ED_{2} \), respectively, are determined by the ED calculated from the flare profile where the normalized flux greater/lower than half of the peak amplitude. 

\begin{figure}
    \centering
    \epsscale{1.2}
    \plotone{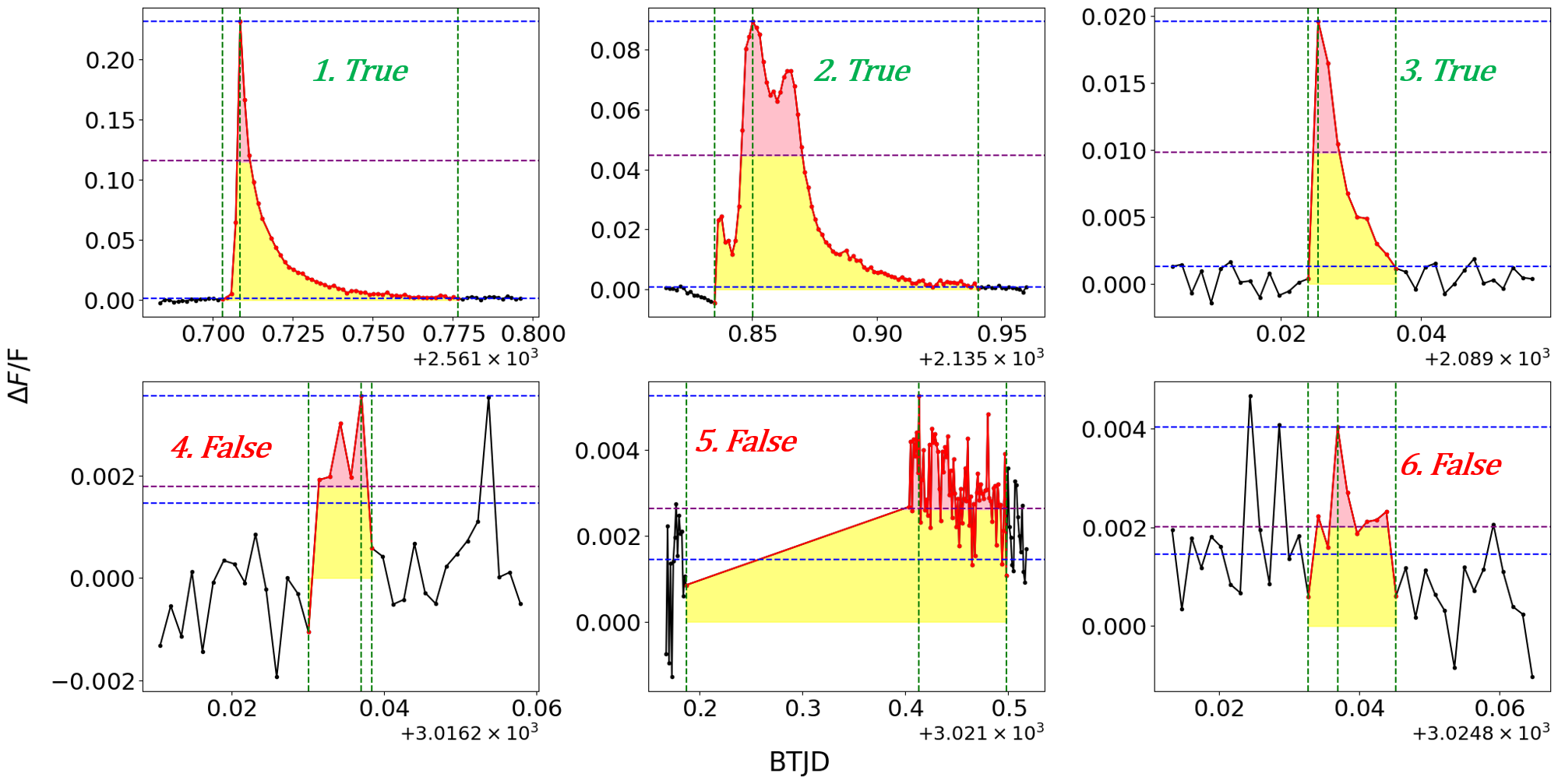}
    \caption{ The example profiles of true (upper three panels) and false (lower three panels) flares. The estimated values of their four classification characteristics are displayed in Tabel~\ref{tab:true_and_false_flare_exp}. The dashed lines and highlighted areas in these panels represent the same properties as described in Figure~\ref{fig:classification_char}.
    }
    \label{fig:true_and_false_flare_exp}
\end{figure}

Figure~\ref{fig:true_and_false_flare_exp} shows the example light curve profiles of “True” flares (upper three panels) and “False” flares (bottom three panels) along with the values of four classification characteristics. 
The true flare can be a single or a multi-event.
True flares tend to have a $\delta t_{1} / \delta t_{2} < 1$ and a larger $A_{f}/\sigma(F_{\text{LC}_{1}})$.  
The duration of true flares is also typically longer than that of false flares.
It is worth noting that the multi-flare with a complex shape could have a $\delta t_{1} / \delta t_{2} > 1$ \citep[e.g.,][]{2022ApJ...926..204H}.
We also included the false sample due to the data gap in the light curve (e.g., the middle panel at the bottom in Figure~\ref{fig:true_and_false_flare_exp}) in our machine-learning models' training.

\subsection{Oversampling the dataset}
\label{subsec:oversampling}
Our current dataset consists of 18,075 samples, but it is imbalanced, with the number of "True" cases (4,589) being significantly lower than the "False" ones (13,486). 
Training a machine learning model on such an imbalanced dataset can lead to biases, potentially skewing the model's accuracy and precision. 
This imbalance may cause the model to favor the majority class ("False" cases in this instance), thus impacting its ability to correctly identify and classify the less represented "True" cases. 
To mitigate these issues and enhance model performance, it's crucial to resolve this imbalance before proceeding with training.

We adopted the Synthetic Minority Oversampling Technique \citep[SMOTE,][]{chawla2002smote}  to generate more synthetic  “True” samples based on the existing ones, introducing balance to our dataset. 
SMOTE does this by randomly selecting an example from the minority class and finding its $k$-nearest neighbors within the same class, where $k=5$ typically. 
It then generates a synthetic sample through interpolation between the selected sample and one of these neighbors. This interpolation process occurs within the feature space (e.g., $\delta t$, $\delta t_{1} / \delta t_{2}$, $ED_{1} / ED_{2}$, and $A_{f}/\sigma(F_{\text{LC}_{1}})$ in this study), effectively expanding the minority class to balance the dataset.
Moreover, compared to the traditional oversampling method, which merely replicates existing cases, this technique enhances model generalization while reducing the likelihood of overfitting by introducing more diverse examples of the minority class. 
{We considered the alternative approach of reducing the sample size of "False" cases to match the number of "True" cases. 
However, this may result in a significant loss of valuable information and potentially lead to underfitting, as the model would have fewer examples to learn from. 
Maintaining a larger dataset by using SMOTE preserves the diversity of the majority class, which is crucial for the model to generalize well to new data. By generating synthetic samples through SMOTE, we retain all the information in the original dataset and improve the model's ability to accurately identify and classify "True" cases.
}

In practice, we applied the SMOTE function from a Python package \texttt{imbalanced-learn} \citep{JMLR:v18:16-365} to augment our original dataset. 
In this way, we generated 8,897 synthetic “True” cases, resulting in a balanced dataset comprising a total of 26,972 samples with half of “True” and half of “False.” 
This balanced, pre-classified dataset will be used for training, validating, and testing our machine learning models.

\begin{figure}
    \centering
    \epsscale{0.7}
    \plotone{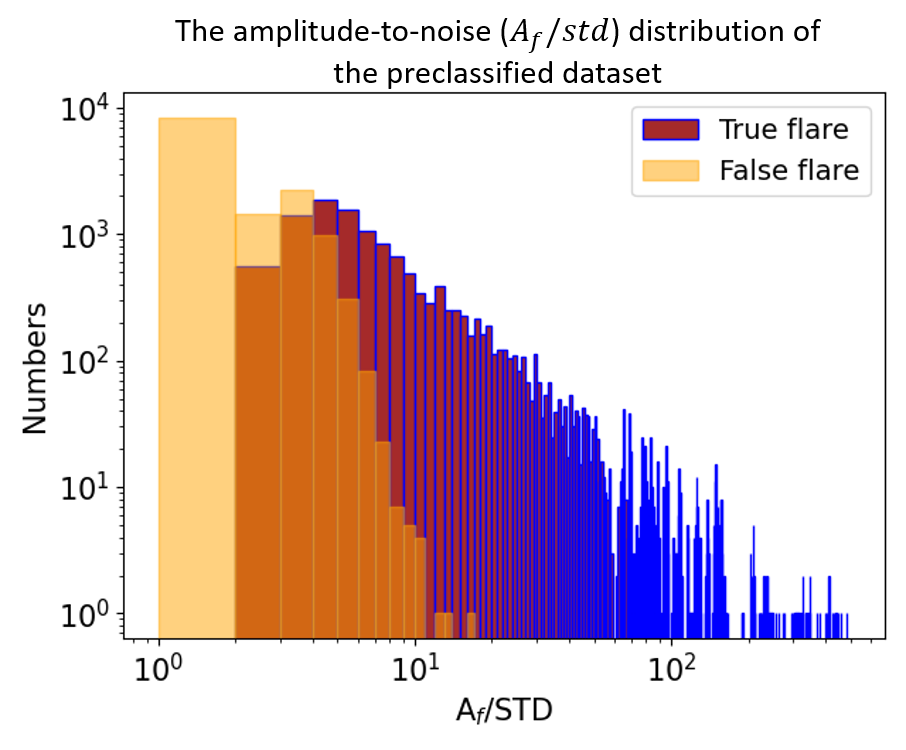}
    \caption{
    The $A_{f}/\sigma(F_{\text{LC}_{1}})$ distributions of the true (purple bar with blue edge) and false flares (orange bar) in our pre-classified dataset. The $A_{f}/\sigma(F_{\text{LC}_{1}})$ of true flares are generally greater than those of false flares and cover a broader range. The number or frequency of flare gradually declines as $A_{f}/\sigma(F_{\text{LC}_{1}})$ increases.
    }
    \label{fig:af-to-std_distribution}
\end{figure}

\begin{figure}
    \centering
    \epsscale{1.2}
    \plotone{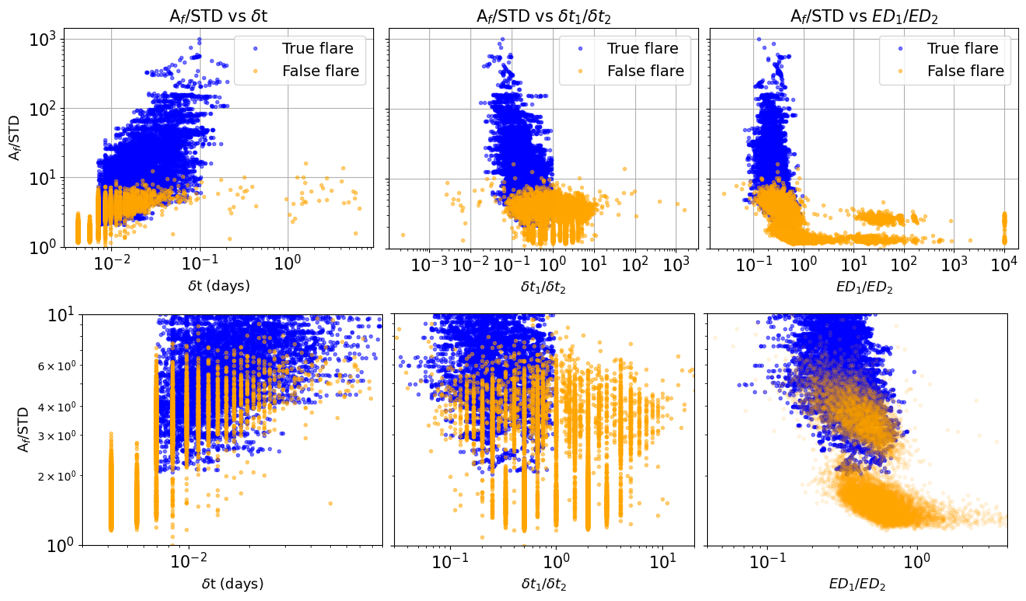}
    \caption{
    The $A_{f}/\sigma(F_{\text{LC}_{1}})$ relationships with $\delta t$ (left panel), $\delta t_{1} / \delta t_{2}$ (middle panel), and $ED_{1} / ED_{2}$ (right panel) in our pre-classified samples dataset.  The distributions of true and false events in the 2-dimensional spaces are also shown. The second column zooms in the area of $A_{f}/\sigma(F_{\text{LC}_{1}})=$1--10.
    For the true flares, their $A_{f}/\sigma(F_{\text{LC}_{1}})$ increase with increasing $\delta t$ but decline as $\delta t_{1} / \delta t_{2}$ grow greater.  The flares with the $A_{f}/\sigma(F_{\text{LC}_{1}}) > 20$ exhibit no apparent trend in $ED_{1} / ED_{2}$, whereas smaller flares could have a larger $ED_{1} / ED_{2}$, yet this ratio does not surpass 1.
    The distributions of all three parameters become less scattering as $A_{f}/\sigma(F_{\text{LC}_{1}})$ grows.  For the false events, the distributions of all three parameters are way more dispersed than those of the true flares. 
    }
    \label{fig:parameter_distribution_vs_af-to-std}
\end{figure}

Figure~\ref{fig:af-to-std_distribution} shows the $A_{f}/\sigma(F_{\text{LC}_{1}})$ distributions of the true and false flares in the dataset.
The difference in the number distribution of true and false flares is significant.
The false events tend to have smaller $A_{f}/\sigma(F_{\text{LC}_{1}})$, whereas true flares demonstrate a broader range. 
It is also noticeable that the true flares typically have a $A_{f}/\sigma(F_{\text{LC}_{1}}) > 2$, and its numbers smoothly decrease with increasing $A_{f}/\sigma(F_{\text{LC}_{1}})$.
Figure~\ref{fig:parameter_distribution_vs_af-to-std} demonstrates the relationships between $A_{f}/\sigma(F_{\text{LC}{1}})$ and the parameters $\delta t$, $\delta t{1} / \delta t_{2}$, and $ED_{1} / ED_{2}$. 
These panels also display the 2-dimensional distributions for both true and false flare events. 
For true flares, there is an observed increase in $A_{f}/\sigma(F_{\text{LC}{1}})$ with a corresponding increase in $\delta t$, while an inverse relationship is found as $\delta t{1} / \delta t_{2}$ becomes larger. 
Flares with an $A_{f}/\sigma(F_{\text{LC}{1}})$ greater than 20 do not show a clear pattern with respect to $ED_{1} / ED_{2}$. 
Smaller flares could have a higher $ED_{1} / ED_{2}$ ratio, and the ratio cannot exceed 1. 
As $A_{f}/\sigma(F_{\text{LC}_{1}})$ increases, the spread in the distribution of these parameters narrows, indicating a more consistent flare profile. 
In contrast, false events exhibit a more dispersed distribution across all three parameters, distinguishing them from the true flares.

\section{Machine Learning Models training}
\label{sec:models_traning}

In the preceding sections, we have prepared all the necessary components for machine-learning. 
Our pre-classified dataset comprises 26,972 samples evenly divided between true flare and false flare classes. 
Each sample is characterized by four features, as defined in Section~\ref{subsec:flare_classification_feature}. 
This section will detail our methodology for training machine-learning models using the pre-classified dataset. 
We will employ three distinct machine-learning algorithms: Deep Neural Network (DNN), Random Forest (RF), and XGBoost. 

{We selected these algorithms due to their unique strengths and suitability for our classification task. 
DNN is highly flexible and can model complex, non-linear relationships, making them ideal for our high-dimensional data \citep{2015Natur.521..436L, 8694781, 2019arXiv190407248B}.
RF, first introduced by \cite{breiman2001random}, is robust in handling large datasets and provides insights into feature importance and capable of performing classification tasks with large datasets with higher dimensionality, ensuring interpretability. 
XGBoost (eXtreme Gradient Boosting) is another ensemble machine learning algorithm \citep{chen2016xgboost}.
It is highly efficient and scalable, offering superior performance and regularization to prevent overfitting.}

{
Other algorithms, such as Support Vector Machines (SVM) and K-Nearest Neighbors (KNN), were considered but deemed less suitable due to their computational intensity and inefficiency with large datasets. Logistic Regression and Naive Bayes were not chosen due to their limitations in capturing non-linear relationships, and Convolutional Neural Networks (CNN) were more complex and better suited for image data rather than our numerical dataset.
}

The training process for each algorithm we employ in this study will be elaborated in three separate subsections below.

\subsection{Deep Neural Network}
We utilized the \texttt{Keras} library within a Python Package \texttt{tensorflow} \citep{tensorflow2015-whitepaper} to construct the architecture of our Deep Neural Networks (DNN) model and train it.
Our DNN model comprises an input layer, a number of hidden layers, and one output layer. 
Each hidden layer contains a certain amount of neuron node units.
Each node processes input data and applies an activation function to produce an output. The activation function introduces non-linearity into the model, allowing it to capture complex relationships in the data. 
In our DNN model, the Rectified Linear Unit activation function \citep[ReLU,][]{2018arXiv180308375A} is used, which is defined as:
\begin{equation}
    \text{ReLU}(x) = \max(0, x),
\end{equation}
where $x$ represents the weighted sum of inputs to the neuron.
For parameter optimization, we adopted adaptive moment estimation \citep[Adam,][]{2014arXiv1412.6980K}, a type of gradient descent algorithm, in our model. 
The output layer produces the final output of the network. 
For classification tasks, softmax activation \citep{NIPS1989_0336dcba} is commonly used to output class probabilities. 
The softmax function for the 
$i$-th class is defined as:
\begin{equation}
    \text{Softmax}(x_i) = \frac{e^{x_i}}{\sum_{j=1}^{K} e^{x_j}},
\end{equation}
where $x_i$ represents the $i$-th element of the input vector $x$, and $K$ is the total number of classes, which in our case is $K=2$, representing "True" and "False" flares. 
The Softmax function computes the exponential of each input value and then normalizes these values by dividing each by the sum of the exponentials of all the input values. 
This produces a vector of values between 0 and 1 that sum to 1, representing a probability distribution over the \( K \) classes.

\begin{figure}
    \centering
    \epsscale{1.2}
    \plotone{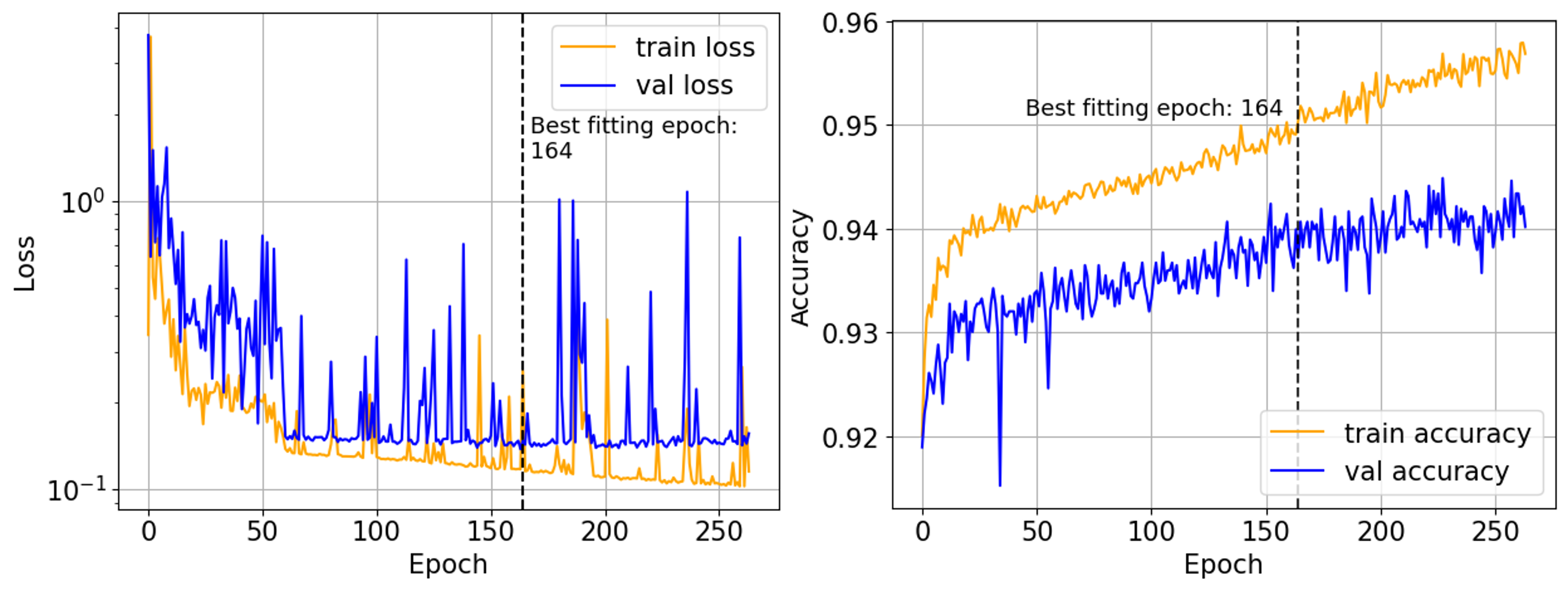}
    \caption{ The loss (left) and the accuracy (right) of the training set and validation set across the training epochs. The black vertical dashed line marks the best-fitting epoch of 164 for our DNN model.
    }
    \label{fig:DNN_best_epoch}
\end{figure}

In order to optimize the hyperparameters for our DNN model and evaluate its performance after being trained completely, we shuffled and divided our pre-classified dataset into three subsets: 70\% for training, 15\% for validation, and 15\% for testing. 
This is the commonly used data splitting ratio, which provides a balance between having enough data for training the model (training set), tuning the model's hyperparameters (validation set), and performing performance evaluation (testing set).
We employed a random cross-validated search algorithm \citep{geron2022hands} for hyperparameters fine-tuning over the predefined ranges of parameters, which are (1) number of hidden layers ($L$: 1–6), (2) number of node units per layer ($N$: 32–512), and the learning rate ($LR$: 0.1, 0.01, 0.001, and 0.0001), while keeping batch size a constant of 32.
The optimal set of hyperparameters for our DNN model, which yielded the highest $F1$~score (see Equation~\ref{eq:F1}, it will be explicitly explained in the next paragraph) during the random search, is listed in Table~\ref{tab:best_hyper_DNN}.
Our model comprises three hidden layers. The first layer has 480 units of nodes, 192 units for the second layer, and 512 units for the third layer.
We then trained our DNN model with these hyperparameters with a 32–batch size for each epoch and a learning rate of $LR=$~0.001.  
To prevent overfitting our model, we monitored the validation loss, calculated using binary cross-entropy \citep{goodfellow2016deep}, and halted training when we observed no further improvement.
The binary cross-entropy loss function, commonly used in binary classification tasks, is defined as:
\begin{equation}
\text{Loss} = -\frac{1}{N} \sum_{i=1}^{N} \Big[ y_i \log(p_i) + (1 - y_i) \log(1 - p_i) \Big].
\label{eq:log_loss}
\end{equation}
In the equation, $N$ is the total sample number defined by the batch size. Thus, in our case, $N=32$ at each epoch.
$y_i$ is the actual class label of the $i$-th sample, and it can take a value of 0 or 1, and
$p_i$ is the predicted probability of the $i$-th sample being of class 1, ranging from 0 to 1.
The term $y_i \log(p_i)$ quantifies the loss for the positive class (when $y_i = 1$).
The term $(1 - y_i) \log(1 - p_i)$ quantifies the loss for the negative class (when $y_i = 0$).
The sum is over all $N$ samples, and the total sum is averaged by dividing by $N$.
Figure~\ref{fig:DNN_best_epoch} displays the progression of loss and accuracy across training epochs for both the training and validation sets.
The optimal model was achieved at epoch 164.

\begin{figure*}
\centering
\gridline{\fig{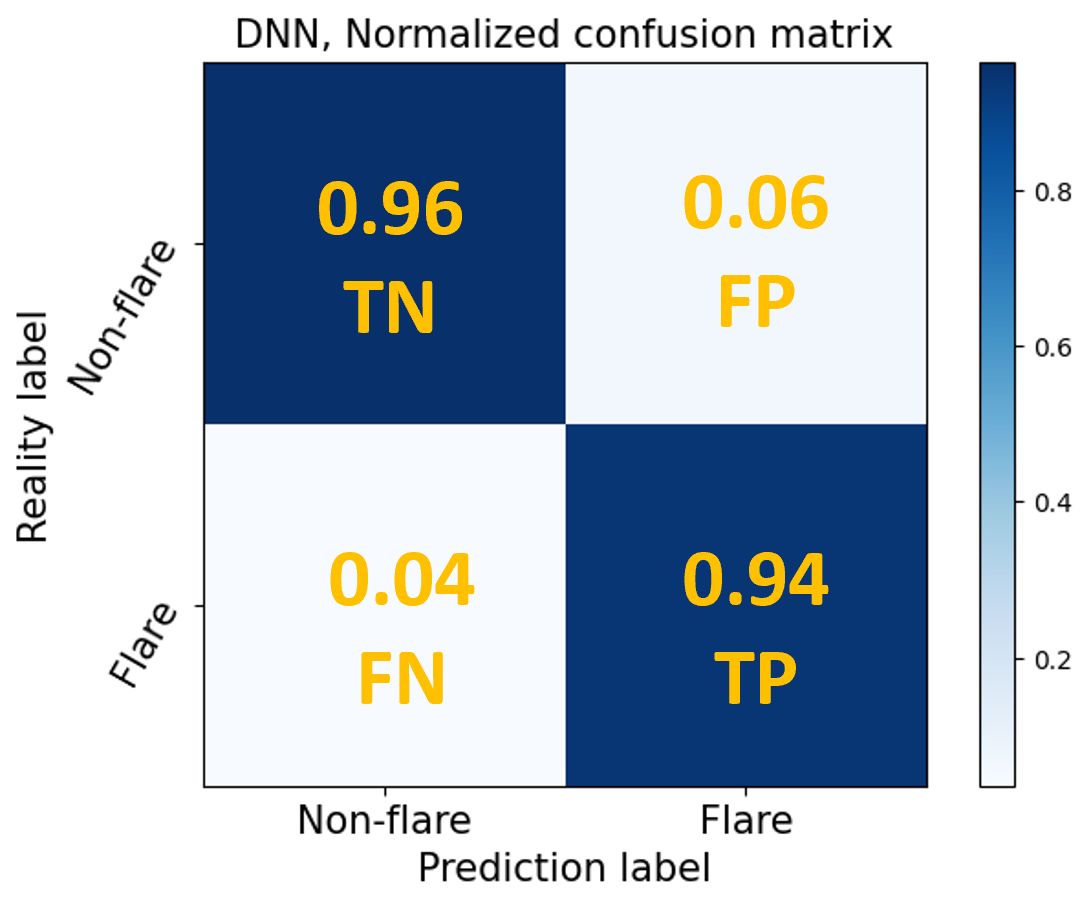}{0.5\textwidth}{(a) Deep Neural Network}
          \fig{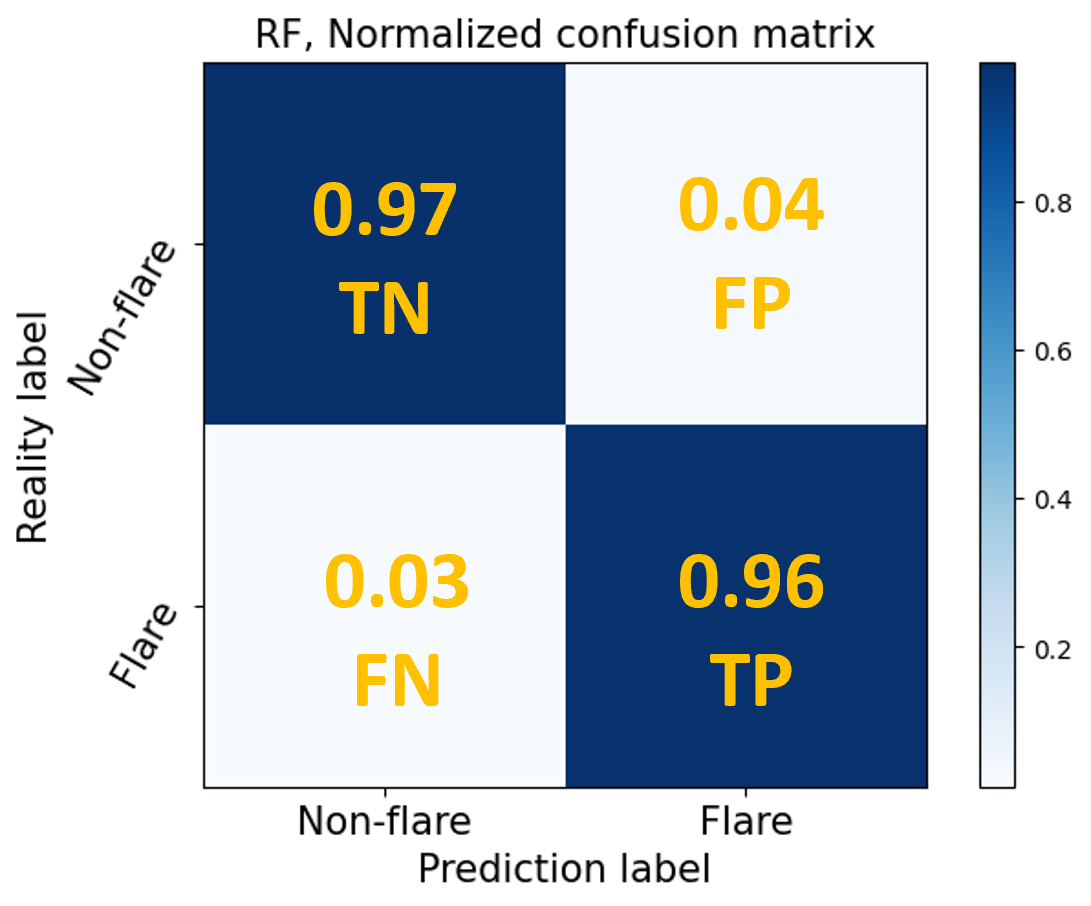}{0.5\textwidth}{(b) Random Forest}
          }
\gridline{\fig{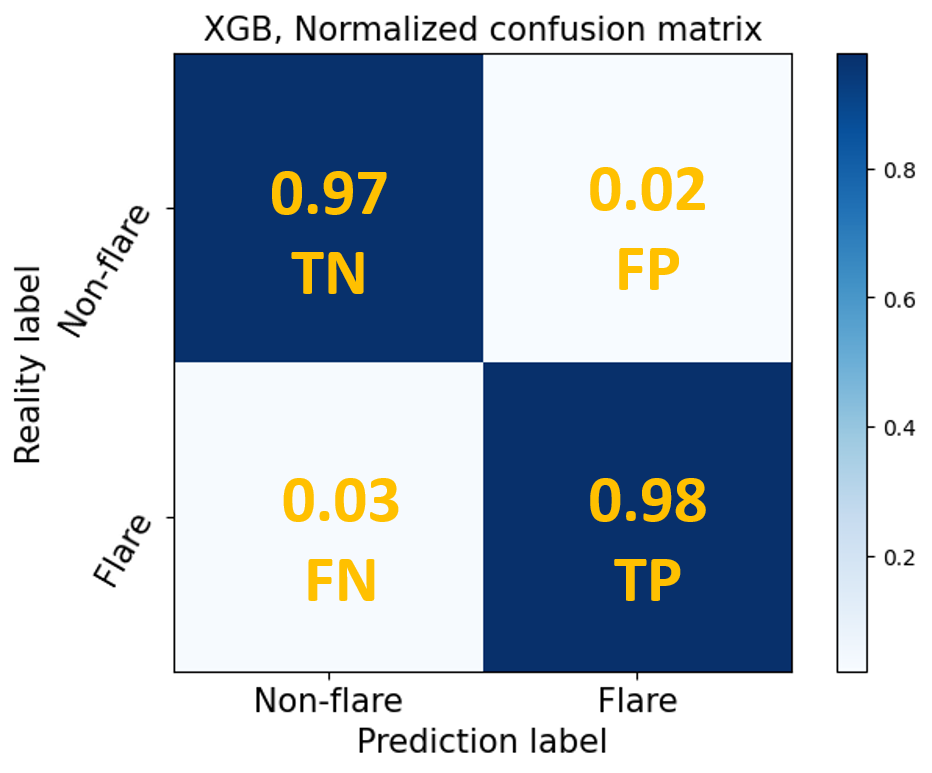}{0.5\textwidth}{(c) XGBoost}
\fig{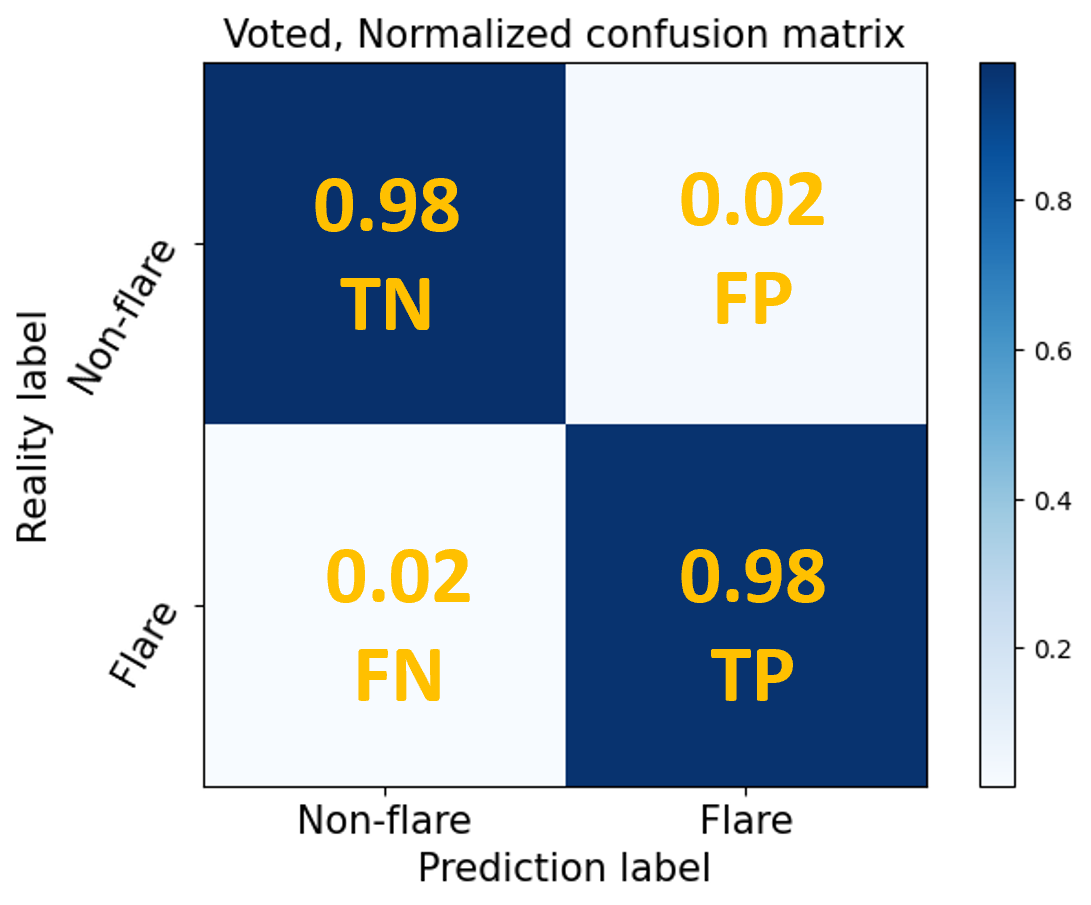}{0.5\textwidth}{(d) Multi-algorithm approach.}
}

\caption{The normalized confusion matrix of the predictions our machine learning models made for the testing set: (a) Deep Neural Network, (b) Random Forest, (c) XGBoost, {and (d) flares voted by all models or multi-algorithm approach.} 
The horizontal label is the predicted class label, and the vertical is the reality class label. The matrix contains the rates of true positive (TP), true negative (TN), false positive (FP), and true negative (TN) of the model's predictions.}
\label{fig:confusion_matrix}
\end{figure*}

We assessed the performance of the trained DNN model with the testing set. 
Figure~\ref{fig:confusion_matrix} shows the normalized confusion matrix of the rates of true positives (TP), true negatives (TN), false positives (FP), and false negatives (FN), which are determined by the predictions of our DNN model made for the testing set. 
We compared four metrics to evaluate these rates, expressed as follows:
\begin{equation}
    Accuracy = (TP+TN) / (TP+TN+FP+FN),
    \label{eq:AC}
\end{equation}
\begin{equation}
    Precision = (TP) / (TP+FP),
    \label{eq:Pr}
\end{equation}
\begin{equation}
    Recall = (TP) / (TP+FN), 
    \label{eq:RC}
\end{equation}
\begin{equation}
    F_{1}~score = 2\times\frac{Precision \times Recall}{Precision + Recall}.
    \label{eq:F1}
\end{equation}
\( Accuracy \) quantifies the proportion of true results (both TP and TN) among the total number of cases examined. \( Precision \) measures the proportion of true positives among all positive results (including both TP and FP), indicating the precision of the positive classifications. \( Recall \), also known as sensitivity, assesses the proportion of actual positives correctly identified, essentially measuring the model's ability to find all relevant cases. Finally, the \( F_{1} \) score considers both precision and recall to provide a balance between the two metrics for a single measure of performance.
The four metrics of our DNN model are listed in Table~\ref{tab:four_metrics_scores}. 
It has an \( Accuracy \) score of 95\%. The \( Precision \) and \( Recall \) are 94\% and 96\%, respectively, giving a \( F_{1} \) score of 95\%.

\subsection{Random Forest}
Random Forest is an ensemble learning technique, which means it relies on the collective decision-making of various predictors to improve overall predictive accuracy and control overfitting. 
The "forest" refers to an ensemble set of decision trees and is usually trained with the bagging method or bootstrap aggregating \citep{breiman1996bagging}.

A decision tree works by breaking down a dataset into smaller and smaller subsets based on the split criteria.
For each node, the split criteria evaluates which one of the classification features that have not been used for splitting along the current branch of the tree is the best feature to split on.
Two primary split criteria are (1) Gini Impurity and (2) Entropy Information Gain.
The Gini Impurity measures the frequency at which any sample in the dataset will be mislabeled when randomly labeled. 
The Gini Impurity for a set \( S \) (a whole dataset for the root node or a split subset for the decision node) is calculated as:
\begin{equation}
       \text{Gini}(S) = 1 - \sum_{i=1}^{n} p_i^2
       \label{eq:gini}
\end{equation}
where \( p_i \) is the proportion of the samples that belong to class \( i \) in the set \( S \), and \( n \) is the number of classes. 
The goal is to minimize this impurity, indicating a more homogenous set of labels. 
Thus, it chooses the feature that reduces the impurity the most to split the node, unless stopping criteria such as the maximum depth of the tree or minimum decrease in impurity are reached.
Entropy, on the other hand, is a measure of the level of disorder or uncertainty in a set. 
In the context of a classification problem, it quantifies the impurity in the dataset. The \( \text{Entropy}(S) \) of a set \( S \) is calculated using the formula:
\begin{equation}
    \text{Entropy}(S) = - \sum_{i=1}^{n} p_i \log_2(p_i)
    \label{ep:entropy}
\end{equation}
where \( p_i \) is the proportion of samples in \( S \) that belong to class \( i \), and \( n \) is the number of different classes.
Information Gain measures the reduction in this entropy resulting from splitting the data according to a particular feature. 
It is used to decide which feature to split on at each step in the tree. The Information Gain \( IG \) from splitting set \( S \) on a feature \( F \) is calculated as:
\begin{equation}
    IG(S, F) = \text{Entropy}(S) - \sum_{v \in \text{Values}(F)} \frac{|S_v|}{|S|} \text{Entropy}(S_v)
    \label{eq:information_gain}
\end{equation}
where \( \text{Values}(F) \) are the different values that feature \( F \) can take, \( S_v \) is the subset of \( S \) for which feature \( F \) has value \( v \), and \( |S_v| \) is the number of samples in \( S_v \).
The feature with the highest \( IG \) is selected for the split, as it represents the feature that provides the most significant reduction in impurity about the target variable.

The nodes stop being split when certain criteria are reached. 
The nodes with no child nodes are referred to as the leaf nodes, and each of them indicates one label (or class) in the classification task.
Eventually, a decision tree with numerous leaf nodes will be built.
When making the prediction for any given input, a decision tree will traverse from the root node to exactly one leaf node, and the prediction for that input is determined by the label associated with that leaf node.
However, the choice of the root node and subsequent nodes can vary greatly depending on the data and the metric used for splitting. 
Different metrics or even different subsets of data can lead to different tree structures. 
This variability is one reason why the ensemble method, i.e., Random Forests, can be more robust and less likely to overfit compared to just using one decision tree.

We used the \texttt{RandomForestClassifier} provided in the \texttt{scikitlearn} Python package \citep{scikit-learn, sklearn_api} to build up the architecture of our random forest model. 
We shuffled 
and split our pre-classified dataset into a training set (75\%) and a testing set (25\%) to constrain the best hyperparameters for our Random Forest model and evaluate the model's performance when training was completed.
The optimal hyperparameters of our Random Forest model were determined from a set of predefined parameter spaces when the random search algorithm found the ones that gave the highest $F1$~score.
These hyperparameters, including their descriptions, are displayed in Table~\ref{tab:best_hyper_RF}.

The confusion matrix of the testing set determined by our Random Forest model is displayed in Figure~\ref{fig:confusion_matrix}.
It nearly outperforms our DNN model in terms of the four metrics.
It achieved an \( Accuracy \) of 97\%. It also attained a \( Precision \)of 96\% and a \( Recall \) of 97\%, resulting in an \( F_{1} \) score of 97\% (Table~\ref{tab:four_metrics_scores}).
An ensemble algorithm like Random Forest can estimate the importance of each input feature in the classification, commonly referred to as "feature importance." 
The feature importance distribution of four input classification characteristics as determined by our Random Forest model are presented in Table~\ref{tab:RF_feature_importance}. 
The feature $A_{f}/\sigma(F_{\text{LC}_{1}})$ has the highest importance weight of 0.58, indicating it has the most significant impact on the model's predictions.
The remaining characteristics $\delta$t, $\delta t_{1} / \delta t_{2}$, and $ED_{1} / ED_{2}$ have  weights of 0.21, 0.14, and 0.07, respectively.

\subsection{XGBoost}
Unlike the Random Forest algorithm, which builds numerous independent trees and combines their predictions, XGBoost is based on gradient boosting, an ensemble technique where new trees are added sequentially to correct the errors made by existing trees until no further improvements can be made. 
This model structure is also generally simpler than that of a Deep Neural Network.
As a result, XGBoost often outperforms Random Forest and DNN models in terms of training and prediction speed.

Specifically, XGBoost starts with an initial prediction, typically the log odds for classification tasks, as expressed by
\begin{equation}
    \text{Log odds} = \log\left(\frac{p}{1-p}\right)
    \label{eq:log_odds}
\end{equation}
where $p$ is the proportion of the positive class in the training set.
This initial prediction is then converted into a probability using the logistic function and substituted into the objective function to process the training. 
During training, XGBoost iteratively adds trees, where each new tree is built to reduce the remaining residuals made by the existing ensemble of trees for the objective function.
The objective function, which guides the training process, is composed of two components: the training loss and a regularization term.
The choice of loss function for XGBoost varies depending on the specific tasks. 
In our case, a binary classification, we employed Eq.~\ref{eq:log_loss}, the same as used in training our DNN model.
Moreover, the criteria used to split nodes within XGBoost trees differs from those used in Random Forest. 
XGBoost uses a more complex criterion that involves a combination of the gradient (the first derivative) and the hessian (the second derivative) of the loss function and a regularization term.
To enhance models' effectiveness and prevent overfitting, XGBoost employs the regularization term to control the complexity of the model.  
This term includes two parts that are similar to the L1 \citep[Lasso Regression,][]{51791361-8fe2-38d5-959f-ae8d048b490d} and L2 \citep[Ridge Regression,][]{a92f3c16-7c6e-31d3-b403-82d2b0a469e4} regularization, respectively.
The overall objective function for a training set with $N$ samples is the sum of the loss function for all samples and the regularization term
\begin{equation}
    \text{Objective Function} = \sum_{i=1}^{N} \text{Loss}_i + \alpha \sum_j |w_j| + \frac{1}{2} \lambda \sum_j w_j^2
\label{eq:OBF}
\end{equation}
where $\alpha \sum_j |w_j|$ is the L1 regularization with $\alpha$ as the regularization parameter, 
$\frac{1}{2} \lambda \sum_j w_j^2$ is the L2 regularization controlled by the parameter $\lambda$, and $w_j$ represents the weight of the $j$-th leaf estimated from both gradient and hessian in the trees \citep{chen2016xgboost}.
In short, XGBoost aims to reduce the objective function's output value by sequentially adding trees until either a minimum is achieved or there is no further improvement.


We used a gradient boosting-based machine learning library \texttt{xgboost}\footnote{https://github.com/dmlc/xgboost. The \texttt{xgboost} is an open-source machine learning library, collaboratively developed and maintained by a community of contributors.} to establish our XGBoost classifier model.
Again, we shuffled and divided our pre-classified dataset into a training set (75\%) and a testing set (25\%). 
This split was essential for fine-tuning the hyperparameters within predefined ranges and evaluating the model's performance upon completion of training. 
For our XGBoost Classifier model, just like with our DNN and Random Forest models, we identified the best hyperparameters within a predefined parameter space by using a random search algorithm. 
This method helped us identify the optimal settings that yielded the highest $F1$~score for the model.
The chosen hyperparameters and their descriptions are detailed in Table~\ref{tab:xgb_hyperparameters}.


The testing set's confusion matrix, as evaluated by our XGBoost model, is shown in Figure~\ref{fig:confusion_matrix}. 
This model surpasses our RF and DNN models in performance based on four key metrics. It achieved an \( \text{Accuracy} \) of 98\%. Additionally, it reached a \( \text{Precision} \) of 98\% and a \( \text{Recall} \) of 97\%, culminating in an \( F_{1} \) score of 98\% (Table~\ref{tab:four_metrics_scores}).
XGBoost is also able to estimate the importance of each input feature in the classification.
The distribution of feature importance for the four input classification characteristics, as determined by our XGBoost model, is outlined in Table~\ref{tab:RF_feature_importance}, along with comparisons to the Random Forest model. 
Notably, the feature \( A_{f}/\sigma(F_{\text{LC}_{1}}) \) has the greatest influence on the model's predictions with a weight of 0.67. 
The remaining features, \( \delta t \), \( \delta t_{1} / \delta t_{2} \), and \( ED_{1} / ED_{2} \), have weights of 0.18, 0.1, and 0.06, respectively.

\section{Model Performance Assessment}
\label{sec:model_performace_assessment}
A key question our study aims to address is the optimal method and its performance for flare identification. 
Therefore, we compare our models against the reference method from \cite{2020AJ....159...60G}, which identified 8,695 single peak flare events in 1,453 TESS 2-minute light curves of 1,228 low-mass stars in Sector~1 and/or 2. 
By conducting a flare search within the same dataset using our machine-learning models, we compare our detection results with theirs.
This comparison will provide additional insights into the effectiveness of our data pre-processing methods and the performance of our machine-learning models.


\subsection{Flare detection by our ML models, the Multi-Algorithm Voting Approach}
We employed the methods and criteria outlined in Sec~\ref{subsec:flare candidates identification} to detrend the long-term variability in the light curves and to select flare candidates, and 113,445 flare candidates were collected. 
We calculated the four classification features for these candidates as described in Sec~\ref{subsec:flare_classification_feature}.
These features were then fed into our machine-learning models for true flare identification.

To mitigate the pseudo-flare-like light curve profiles caused by sub-optimal de-trending models and enhance the accuracy of flare classification, the flare must also satisfy the following criteria: (1) A valid flare event must not occur near any gaps in the data. Specifically, the absolute time difference between the peak of the flare and the nearest data gap must be longer than 1200~second.
(2) The duration of the impulsive phase must be shorter than the decay phase, i.e.,$\delta t_{1} / \delta t_{2}<1$. (3) The equivalent duration above the half-peak amplitude of the light curve profile must be smaller than below half-peak, i.e., $ED_{1} / ED_{2}<1$.  
These criteria are designed to exclude false positives that may arise from data irregularities by referring to the criteria described by \cite{2020A&A...637A..22R} and what we have observed in Figure~\ref{fig:parameter_distribution_vs_af-to-std}. 
We acknowledge that the criteria above might exclude some complex flare events, particularly multi-flares that consist of a smaller flare followed by a larger one during the decay phase of the initial flare. 
We recognize, however, that the identification of such rare events can be of special interest that we will revisit in future work.

The application of these criteria combined with our DNN model yielded 8,391 flare events, while the XGboost model found 7,708 events.
The Random Forest model recognized 8,951 candidates as true events, substantially higher than that of the other two models. 
This difference led us to conduct a detailed examination of the outcomes produced by our models, involving a visual review of flare profiles within randomly selected light curves.
We discovered that the Random Forest model is more sensitive than others to small bumps with a \( A_{f}/\sigma(F_{\text{LC}_{1}}) \) between 2 and 3.
Some of these RF-identified events are too subtle for the DNN and XGBoost models to recognize.

{To ensure more reliable detection, we subsequently introduced a multi-algorithm voting approach, defining flares recognized by at least two different machine learning models ("voted flare" hereafter) as the most probable true flares. 
To address potential concerns regarding the out-voting issue, i.e., one model significantly outperforms the others and the others potentially out-vote or misclassify true events, we conducted a performance analysis using the same testing set as used for evaluating the DNN and XGBoost models.
The results, shown in Figure~\ref{fig:confusion_matrix} (panel d) for the confusion matrix and in Table~\ref{tab:four_metrics_scores} for the performance metrics, demonstrate that the multi-algorithm approach achieves a balance between the three models. 
The overall performance of the multi-algorithm approach is better than that of each individual model. 
While some true flares might be out-voted, the fraction of these events is extremely small, as indicated by the minimal difference in precision between the multi-algorithm approach and the XGBoost model. 
Consequently, the influence of the potential out-voting issue can be considered negligible in the subsequent statistical analysis in our study.}

\begin{figure}
    \centering
    \epsscale{1.2}
    \plotone{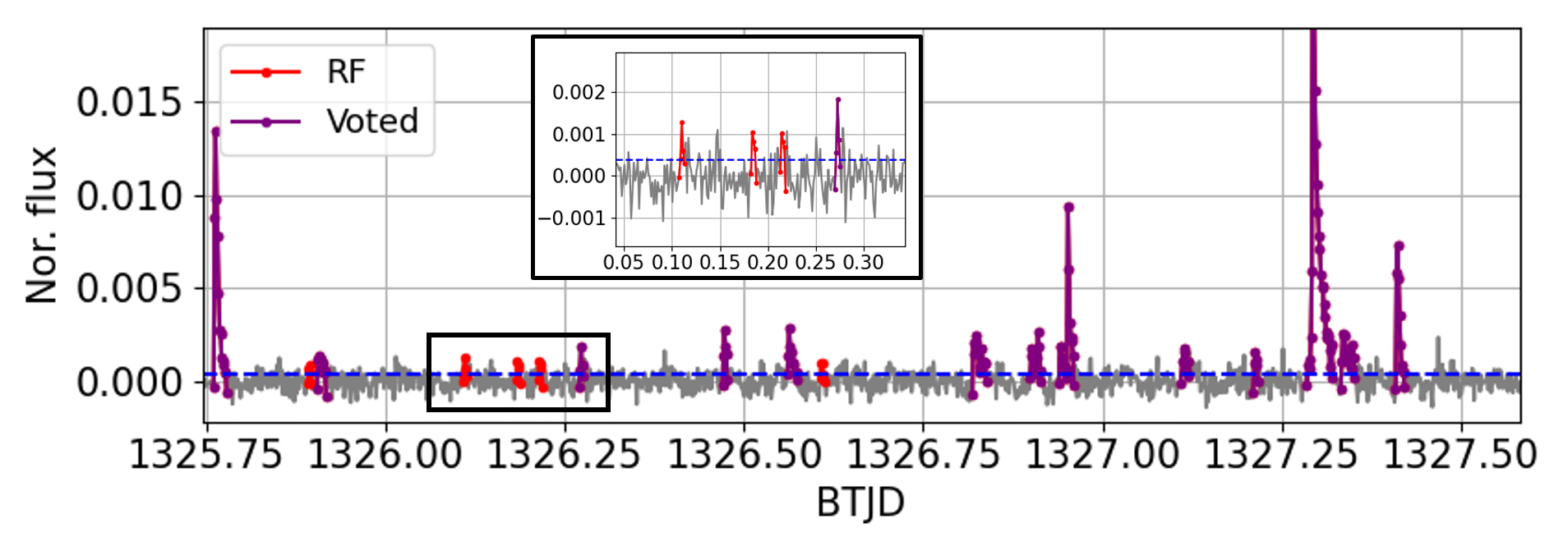}
    \caption{A partial 2-min light curve of TIC~140045538, an extremely flare active star reported by \cite{2020AJ....159...60G}. The light curve profiles in red and purple are the flares detected only by the Random Forest model and voted flares, respectively.  
    The blue horizontal dashed line represents the 1-$\sigma$ level of noise of the flattened light curve, i.e., \(\sigma(F_{\text{LC}_{1}}) \).
    RF-only flares are generally smaller than voted flares. 
    The inset panel in the figure shows the detailed view of the pattern enclosed within a black box between BTJD~=~1326.05 and 1326.35. This highlights the profile of RF-only flares and offers a detailed comparison with small voted flares.
    }
    \label{fig:flare_detection_comparison_RF-vs-voted}
\end{figure}

\begin{figure}
    \centering
    \epsscale{0.55}
    \plotone{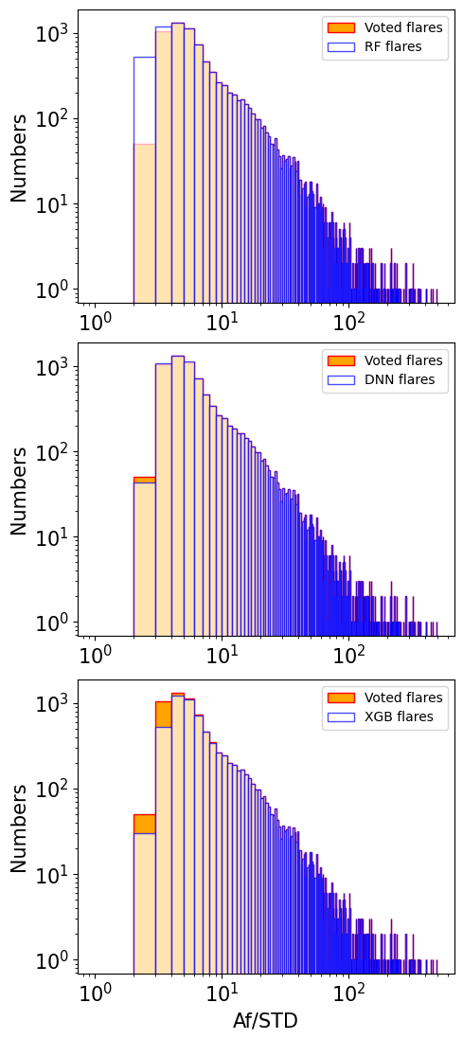}
    \caption{
    A series of comparisons of \( A_{f}/\sigma(F_{\text{LC}_{1}}) \) distributions between the RF (top), DNN (middle), XGBoost (bottom), and voted flares. The Random Forest model reported significantly more small flare with \( A_{f}/\sigma(F_{\text{LC}_{1}}) \) between 2 and 3 than the others.
    }
    \label{fig:flare_detection_comparison_hist_rf_dnn_xgb-vs-voted}
\end{figure}

Figure~\ref{fig:flare_detection_comparison_RF-vs-voted} illustrates this phenomenon by displaying a partial 2-minute light curve of TIC~140045538, an extremely flare active star in this test with 157 voted flares during Sector 1. 
The figure shows that these RF-only flares are generally smaller than the voted flares.
Figure~\ref{fig:flare_detection_comparison_hist_rf_dnn_xgb-vs-voted} shows a series of comparisons of \( A_{f}/\sigma(F_{\text{LC}_{1}}) \) distributions between the RF, DNN, XGBoost, and voted flares. 
The Random Forest method reported a significantly greater number of small flares with a \( A_{f}/\sigma(F_{\text{LC}_{1}}) \) between 2 and 3.
As a result, the count of "voted flares" is 8,357 in total.

\subsection{Comparison to \cite{2020AJ....159...60G}}
\begin{figure}
    \centering
\gridline{\fig{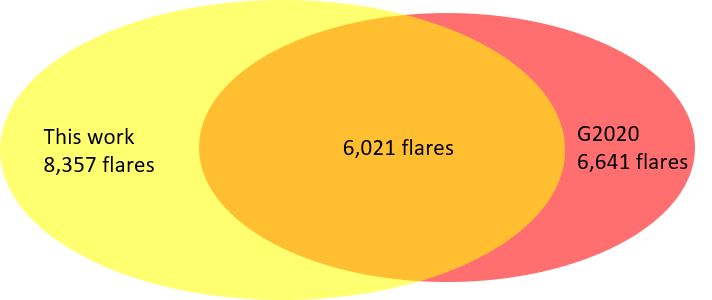}{0.65\textwidth}{(a)}}
\gridline{\fig{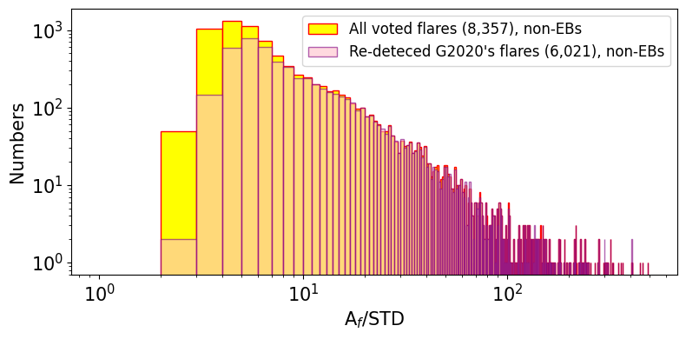}{0.7\textwidth}{(b)}}
    \caption{(a) A Venn diagram illustrates the flare detected from the same TESS dataset by \cite{2020AJ....159...60G} (red) and our models (yellow) and the detection overlap. (b) The \( A_{f}/\sigma(F_{\text{LC}_{1}}) \)  distributions of all voted flares (the flares recognized by at least two different ML models) and those also reported by \cite{2020AJ....159...60G}.
    }
    \label{fig:amp-to-std_all_voted_and_g2020}
\end{figure}

We further checked the number of flare events detected by both \cite{2020AJ....159...60G} and our machine learning models. 
It is important to note that while our models are capable of identifying multi-flare events as true events, they lack the capability to decompose such events into the component single flares. 
Numerous flares reported by \cite{2020AJ....159...60G} were identified as component flares within a multi-flare event or as part of an 'outburst epoch,' according to their definition. 
Conversely, an outburst may consist of one single flare event. 
In other words, our machine learning models detect only outburst epochs as \cite{2020AJ....159...60G} 
defined.
In this study, for convenience, we refer to an outburst epoch as a flare event, which may be a multi-flare or a single-peak flare.
Furthermore, given that our generalized approach to light curve detrending might not be effective for eclipsing binary systems (a limitation that will be thoroughly addressed in the subsequent section), we have excluded these targets from our comparative analysis.
Under this condition, the number of flares from \cite{2020AJ....159...60G} is reduced to 6,641.
Our machine learning models successfully recognize $>90\%$ of these events. 
Specifically, the DNN model found 6,239 flares ($94\%$), the Random Forest model identified 6,265 flares ($94\%$), and the XGBoost classifier detected 6,140 flares ($92\%$).
Overall, approximately $91\%$ (6,021) of the events are the voted flares (Figure~\ref{fig:amp-to-std_all_voted_and_g2020}-a).
We expected to recover about $96\%$ of the events based on the recall scores of our models.
The lower-than-expected recall rates may be attributed to various reasons, such as the different light curve detrending methods we applied.
Besides the re-detected flares, we have found 2,336 more events than \cite{2020AJ....159...60G} presented.
Figure~\ref{fig:amp-to-std_all_voted_and_g2020}-b shows the \( A_{f}/\sigma(F_{\text{LC}_{1}}) \)  distributions of all voted flares and those also detected by \cite{2020AJ....159...60G}.
Apparently, our models are generally much more sensitive to the small flares than the flare detection approach of \cite{2020AJ....159...60G}.
Only four events from them have \( A_{f}/\sigma(F_{\text{LC}_{1}}) \)$~\leq 3$.
Notably, 8,307 of our voted flares have an \( A_{f}/\sigma(F_{\text{LC}_{1}}) \)~$\geq$~3. 
This still substantially exceeds the number of detections reported by \cite{2020AJ....159...60G}.

\begin{figure}
    \centering
    \epsscale{1.1}
    \plotone{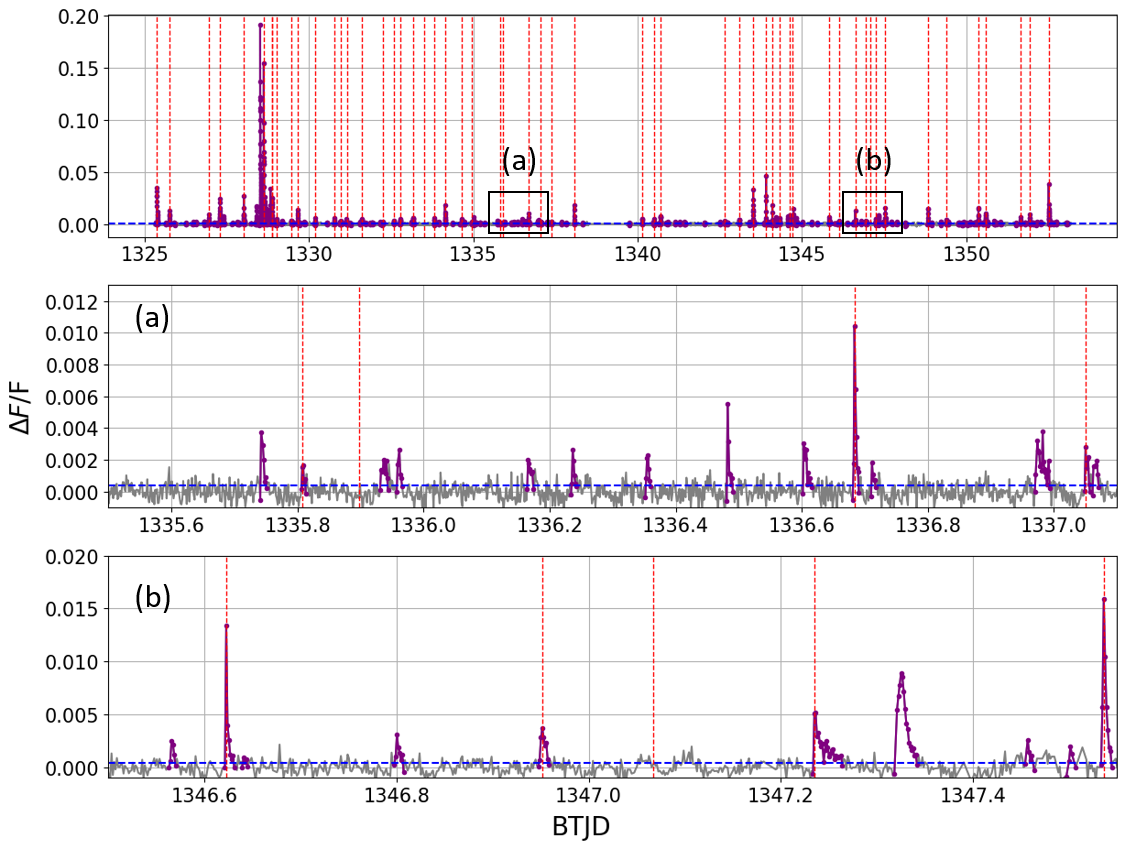}
    \caption{The 2-min TESS light curve of TIC~140045538 observed in Sector 1 (top panel). 
    The blue horizontal dashed line represents the 1-$\sigma$ level of noise of the flattened light curve, i.e., \(\sigma(F_{\text{LC}_{1}}) \).
    In total, the purple profiles present 158 voted flares detected by our ML models. 
    The red vertical dashed lines mark the time position of 56 flares reported by \cite{2020AJ....159...60G}.
    The middle and bottom panels display the zoomed-in views of the time spans in box~(a) and box~(b) from the all Sector~1 time span, respectively.
    }
    \label{fig:flare_profiles_in_light_curve_g2020-vs-ours}
\end{figure}

\begin{figure}
    \centering
    \epsscale{1.1}
    \plotone{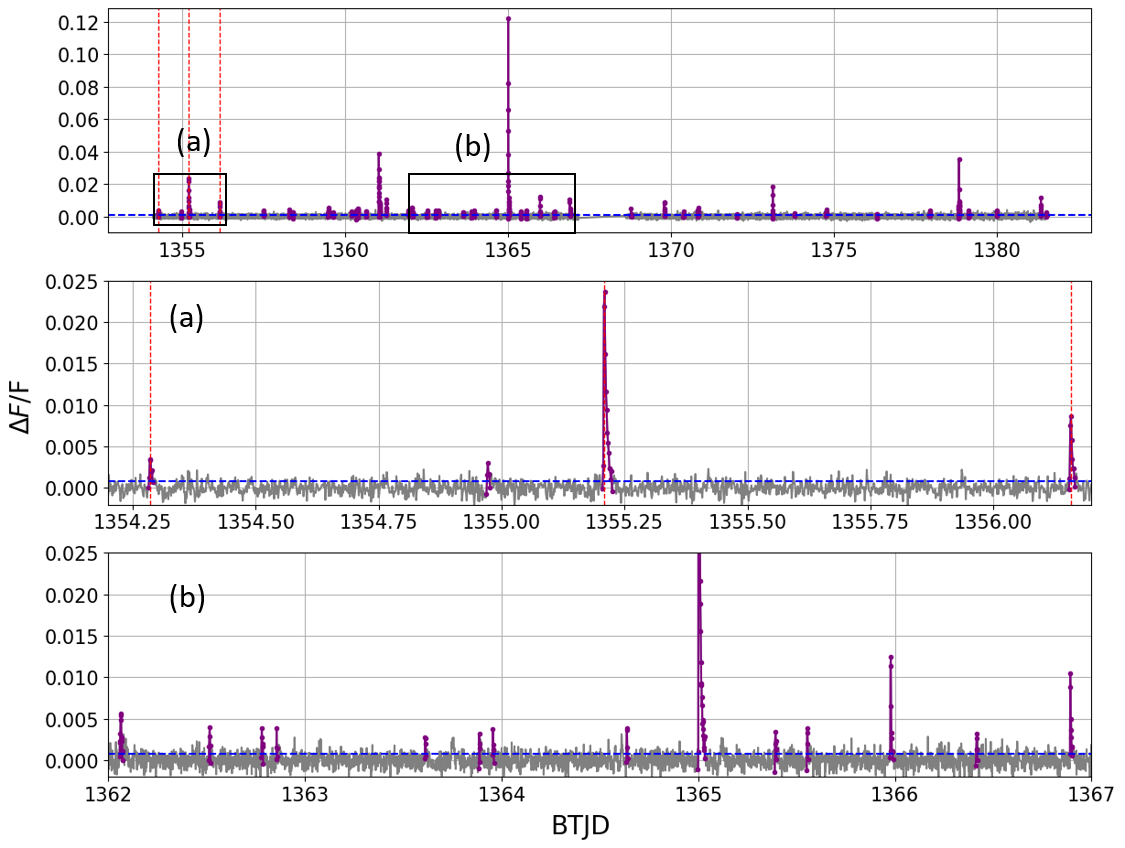}
    \caption{The 2-min TESS light curve of TIC~434103039 observed in Sector 2 (top panel). 
    The horizontal dashed line in blue represents the 1-$\sigma$ noise level of the flattened light curve.
    The light curve profiles in purple are 40 voted flares detected by our ML models. 
    The red vertical dashed lines represent the time position of the flares from \cite{2020AJ....159...60G}.
    The zoomed-in time spans in box~(a) and box~(b) from the top panel are shown in the middle and bottom, respectively.
    }
    \label{fig:flare_profiles_in_light_curve_g2020-vs-ours-tic434103039}
\end{figure}

To illustrate the reality of the flares identified by our models and to conduct a detailed comparison with the detections reported by \cite{2020AJ....159...60G} in terms of flare profiles in light curves, we show two stars as examples: TIC~140045538 and TIC~434103039.
We identified 158 voted flares in the Sector~1 2-minute light curve of TIC~140045538, significantly greater than 56 outburst epochs \cite{2020AJ....159...60G} reported (see Figure~\ref{fig:flare_profiles_in_light_curve_g2020-vs-ours}).
\cite{2020AJ....159...60G} only detected three flares during the TESS Sector~2 observation of the star TIC~434103039 at the beginning; they somehow missed all promising flares after BTJD~$=$~1357 in the light curve.
In contrast, our models identified 40 flares in the light curve (see Figure~\ref{fig:flare_profiles_in_light_curve_g2020-vs-ours-tic434103039}).
Consequently, our models have proven their capability at high sensitivity and are fully functional for conducting flare identifications in TESS 2-minute light curves.

\subsection{Limitations of generalized detrending method and segmentation strategy for improving flare detection reliability}
\label{sec:rotation_period}
During our testing phase with \cite{2020AJ....159...60G} dataset, we have encountered limitations with our current pre-processing methods, particularly in accurate light curve de-trending for eclipsing binaries and stars with a rapid rotation period and/or a substantial modulation amplitude.
This issue often results in the inability to identify flare candidates in the preprocessing phase and to accurately classify true flare events in the machine learning models' tasks for such targets.
The above issue is similar to the issue that has been extensively discussed by \cite{2023A+A...669A..15Y}.
Given our objective to extend flare detection across the entire TESS short-cadence dataset, achieving an ideal de-trending approach for every variety of light curve, without adjusting specific parameters, is not feasible. 
Therefore, we propose grouping the stars into four quadrants, categorized by variation amplitude and rotation period. 
{Our goal is to determine the range of rotation periods and variation amplitudes for which our preprocessing method can effectively detrend the light curves, and we will focus on the stars within these ranges (i.e., in the particular quadrant) in this study.}

\begin{figure}
    \centering
    \epsscale{1.1}
    \plotone{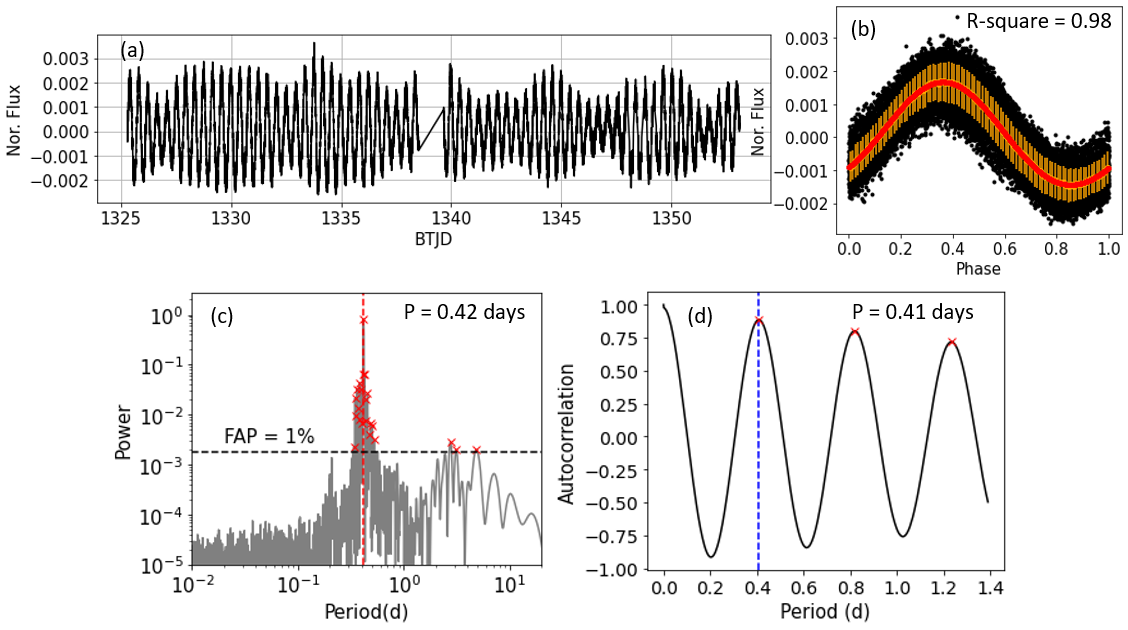}
    \caption{
    The demonstration outlines our approach to estimating the rotation period of the stars.
    The panel (a) shows a short-cadence light curve of TIC~126803030 observed in the TESS Sector 1.
    The panel (b) shows the phase curve (black) created by folding the light curve with a period of 0.42~days.
    The binned phase curve (yellow) is produced by averaging the normalized flux and the corresponding one standard deviation within the bins with a bin width of 0.005 phase.
    We fit the binned phase curve with a 5th-order Fourier series function (red curve) and calculate the the R-square of the fit model to evaluate the validity of the period. 
    In this case, the R-square of the fit is 0.98.
    This period is estimated from the Lomb-Scargle periodogram, shown in the panel (c).
    The horizontal black dashed line represents the false alarm probability (FAP) threshold of 1\%. 
    The peaks with a FAP smaller than the threshold are marked with the red crosses. 
    The vertical red dashed line pinpoints the highest peak in the periodogram, which corresponds to a period of 0.42~days.
    The panel (d) is the Autocorrelation Function of the light curve. 
    The vertical blue dashed line represents the best period of 0.41~days from the ACF, which is consistent with the result derived from the LS method with a $<5\%$ divergence.
    Given the high R-square value and the agreement between the LS and ACF methods, this period is considered a valid measurement.
    }
    \label{fig:LS-period-estimate-tic126803030}
\end{figure}

We started from establishing a metric to quantify the variation amplitude of a star, which we called the variation amplitude index (V$_{amp}$). 
We first masked out the data point with a flux greater than 3-$\sigma$ of the raw normalized pdcsap light curve. Then we define the V$_{amp}$ as the 3-$\sigma$ flux level of the remaining data points.
This ensures that the index is not affected by the outliers, such as flares and transits due to stellar or planetary companions, too severely and is able to reflect the intrinsic variability of the star.
For the rotation period, we used both the Generalized Lomb-Scargle periodogram (GLS) \citep{2009A&A...496..577Z}  from the Python package \texttt{astropy.timeseries} \citep{2013A&A...558A..33A} and the Autocorrelation Function (ACF) to estimate this parameter for every star.
We selected the highest peak with a false alarm probability lower than 1\% in the periodogram as the LS rotation period of the star. 
We then produced the phase curve by folding the raw light curve with this period and examined the validity of the period by fitting the binned phase curve with a 5th-order Fourier series function, estimating the R-square value between the best-fit model and the binned phase curve. 
We employed the ACF to further validate the period derived from the LS method.
The ACF~period is the value at the most significant peak in the auto-correlation function.
For a period measurement to be considered valid, it must satisfy two criteria: 
(1) the measurement must yield an R-square value exceeding 0.7; (2) the period must be consistently identified by both the Lomb-Scargle and the ACF analyses, with the stipulation that the periods derived from both methods do not diverge by more than 5\%.
Figure~\ref{fig:LS-period-estimate-tic126803030} demonstrates the entire process of our period estimation, taking a star TIC~126803030 in Sector~1 as an example, with a rotation period of 0.42~days.

\begin{figure}
    \centering
    \epsscale{0.8}
    \plotone{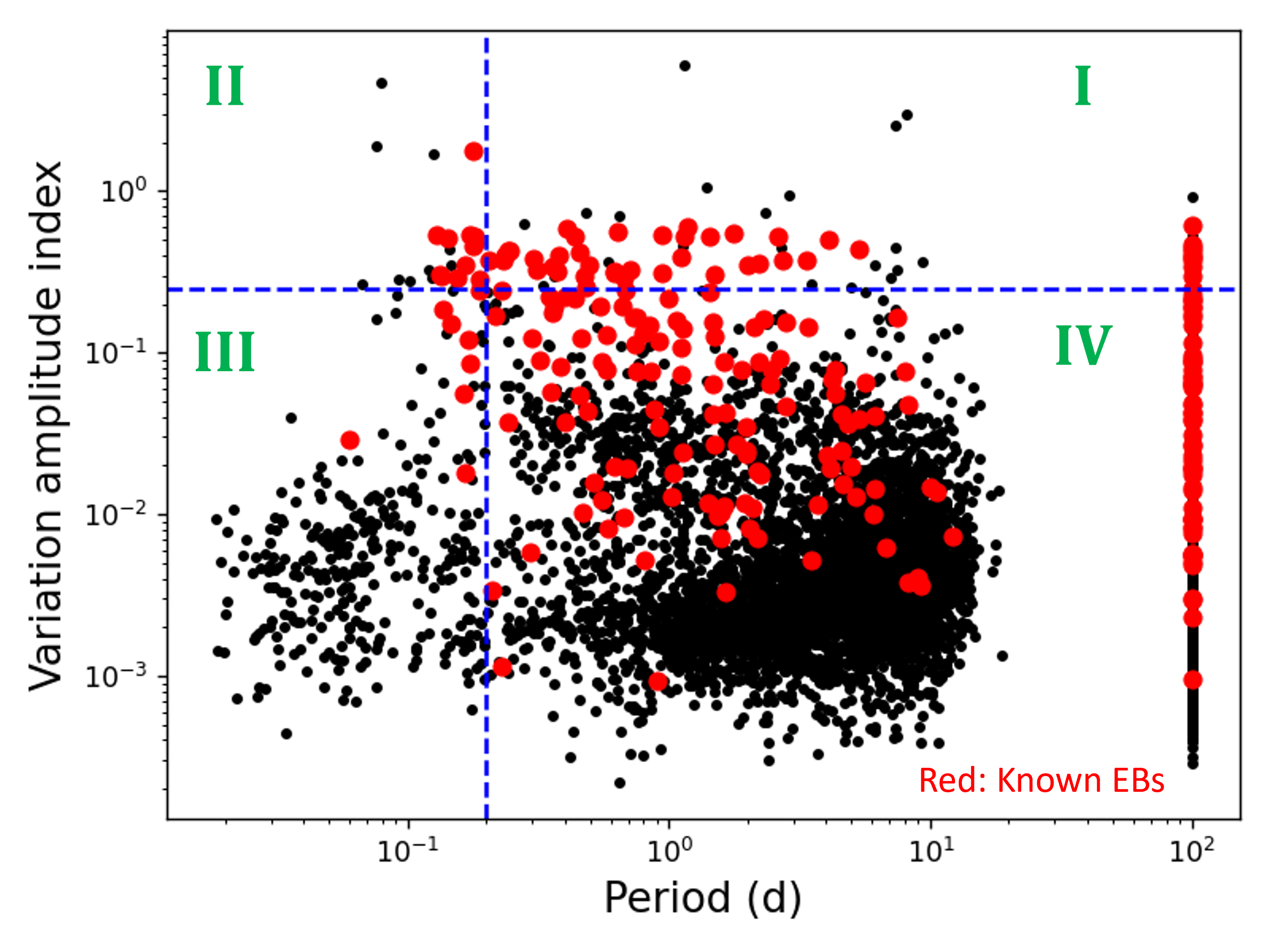}
    \caption{The Period-V$_{amp}$ diagram of the short-cadence stars in the TESS Sector~1.
    The red circles are the eclipsing binaries identified by \cite{2022yCat..22580016P}.
    For the stars with no valid rotation period in this study, we assigned to them a default period of 100 days for simplicity and ease of data handling. 
    Note that this does not reflect the actual periods of the objects in question.
    The horizontal and vertical blue dashed line represent the benchmarks of rotation period (P$_{rot}$~=~0.2~days) and variation amplitude index (V$_{amp}$~=~0.25), respectively.
    Based on these, the stars are divided into four quadrants (numbers are highlighted in green color).
    Excluding eclipsing binaries, approximately 95\% of stars are located in Quadrant~IV.
    These stars are generally fitted well by our detrending method, and are, therefore, the primary focus of this study.
    }
    \label{fig:four-quadrants_sector1}
\end{figure}

We executed the pre-processing phase on the entire dataset from Sector 1 2-minute observations, for which we calculated the variation amplitude index (V$_{amp}$) and rotation period. 
Upon visually inspecting various light curves, we verified that our de-trending method generally performs well for stars that have a rotation period longer than 0.2 days and a V$_{amp}$ less than 0.25, and vice versa.
The stars having no valid rotation period due to the limited observation time-span and a V$_{amp}$~$<$~0.25 are also generally fitted well by our de-trending method.
Therefore, we categorized the stars into four quadrants in the Period-vs.-V$_{amp}$ diagram based on these thresholds (Figure~\ref{fig:four-quadrants_sector1}, for the stars in the Sector~1).
Notably, stars that are well-fitted by our detrending approach predominantly fall into Quadrant IV of the diagram, which contains 95\% of the total stars.
Stars located in Quadrants I and III appear to be less compatible with our detrending method, suggesting that one should be cautious when interpreting the flare detection results of these stars. 
Stars in Quadrant II, in particular, demonstrate a significant mismatch with our method, leading us to advise against relying on flare detection results from these stars.
Additionally, eclipsing binaries located in any quadrant may also experience issues due to a poor fit by our detrending method.
For the stars in Quadrants I, II, III, and eclipsing binaries, there is a clear need for a more generalized detrending algorithm, which we would like to address in future work.

\section{Implementation across the full dataset of TESS short-cadence observations}
\label{sec:deployment_on_tess_data}
By 2024, Jan. 1, there are 72 sectors (Sector 1 to Sector 72) of TESS 2-minute data published, comprising a total of 1,286,260 light curves from 494,326 sources. 
Our goal is to apply our machine learning models to this extensive dataset to develop a comprehensive catalog of flare stars observed in TESS short-cadence data. 

We utilized the High Performance Computing (HPC) system if the University of Arizona to execute the flare searching tasks across the entire TESS Sector~1--72 2-minute dataset. 
These tasks, including light curve detrending, flare candidates selection, and flare identification, were accomplished by using eight CPU processors and 32~GB random access memory in approximately 300 CPU hours.
Additionally, the computing task for variation amplitude index and rotation period takes another 400 CPU hours to complete. 

\subsection{Flare detection results}
Our models identified 18,032 quadrant IV flare stars with a total 249,562  flares in the entire TESS 2-minute light curves from Sector~1 to Sector~72.
We note that none of these stars is currently recognized as an eclipsing binary (EB) in the literature; however, it is possible that some unknown EBs might be included.
{Out of 18,032 flaring stars, 12,253 have the R$_{cont}$ value $<$ 0.1. We detected 173,709 flares in these stars, indicating that 70\% of the observed flares are unlikely to be contaminated. Conversely, 75,833 or 30\% of observed flares were detected from the remaining 5,779 stars with R$_{cont}$ $>$ 0.1.}

\begin{figure}
    \centering
    \epsscale{0.7}
    \plotone{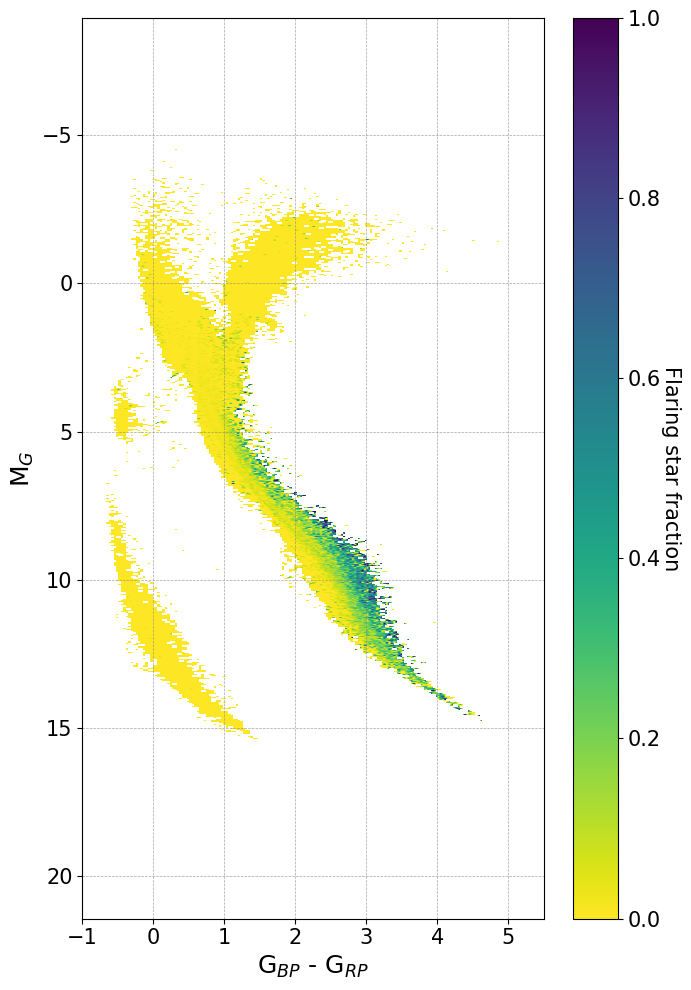}
    \caption{The color-magnitude diagram flare star fraction map. 
    The horizontal axis represent the color index estimated from Gaia~BP and RP mag.
    The vertical axis shows the absolute Gaia's G~mag.
    The gradation of colors, transitioning from a sparser distribution of {yellow to a more concentrated region of dark blue}, signifies the increasing fraction of stars with detected flaring phenomena in the TESS 2-min survey from Sector~1 to Sector~72.
    }
    \label{fig:flaring_fraction_CMD}
\end{figure}

Figure~\ref{fig:flaring_fraction_CMD} presents a detailed color-magnitude diagram ($BP-RP$ color versus $M_{G}$ in the Gaia system) illustrating the distribution of flare star fractions across various stellar populations. 
{The plot's color density, ranging from yellow to dense dark region}, illustrates the increasing fraction of stars exhibiting flaring activity during the TESS 2-minute observation from Sector~1 to Sector~72. 
A large number of the stars that show flares are located along the main sequence.
Notably, there is a marked increment in flaring activity among stars with a $BP-RP$ color index ranging from 1 to 2, corresponding to spectral types K0 to K9.
The regions exhibiting the highest intensity of flaring activity are found for $BP-RP$ color less than 2, highlighting M~dwarfs as a particularly active group.

\begin{figure}
    \centering
    \epsscale{1.2}
    \plotone{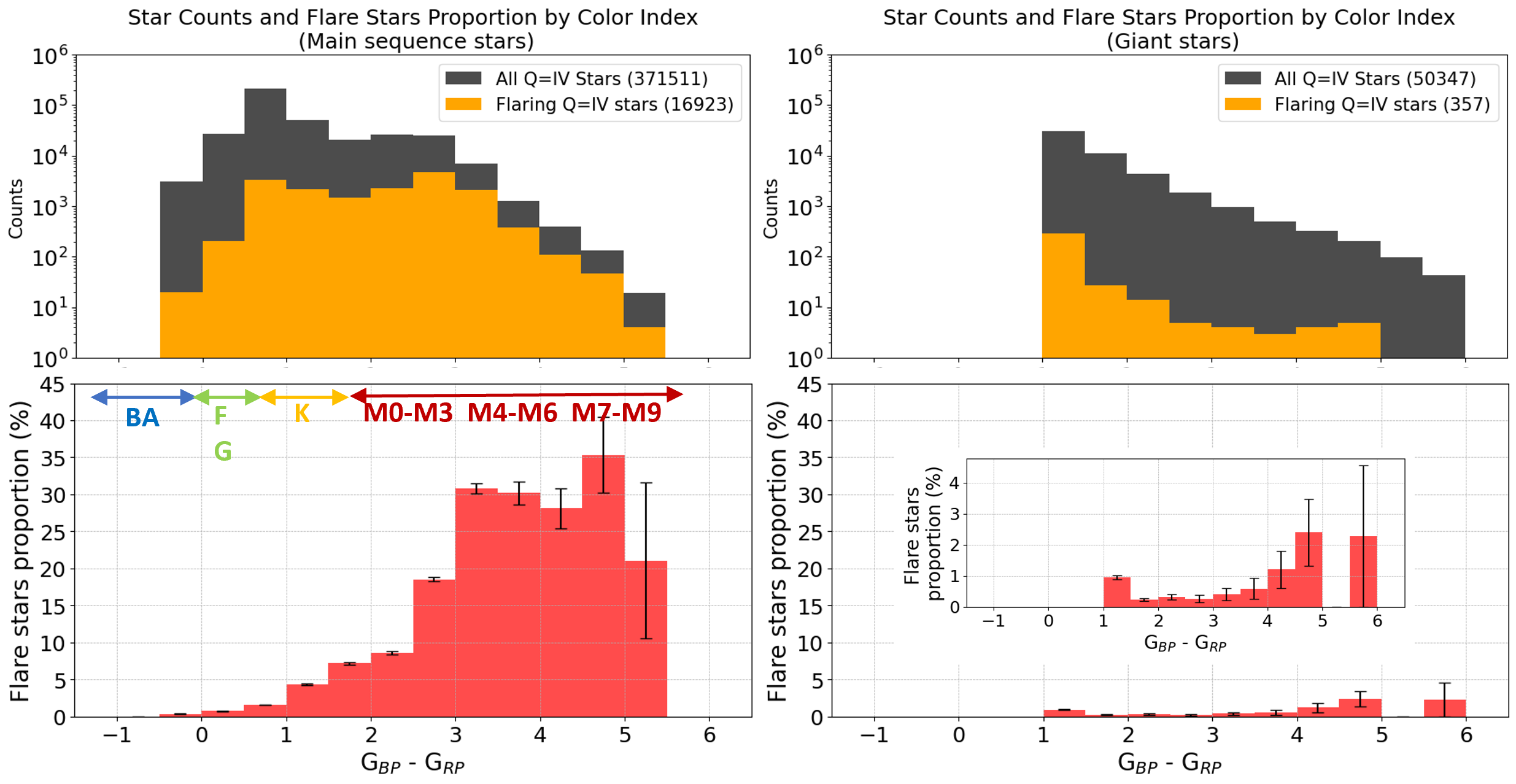}
    \caption{
    Distribution of star counts and flare star proportions by color index.
    {All stars in this figure are quadrant IV stars.}
    The top panels show the total number of stars (gray) and flaring stars (orange) for each color index. 
    The bottom panels illustrate the proportion of flaring stars with error bars.
    {We identified 16,923 flare stars out of 371,551 main sequence stars, indicating that about 4.6\% exhibit flare activity.
    The proportion of flare stars increases with redder color indices, peaking at $BP-RP$ values between 3 and 3.5, corresponding to spectral types M4 and M5.
    Beyond M5-M6, the proportion of flare stars decreases but then shows an increase at M7.
    Among giant stars, which include subgiants and red giants, approximately 0.7\% (357 out of 50,347 stars) exhibit flaring activity}
    }
    \label{fig:flare_stars_proportion_histogram}
\end{figure}

{We further analyzed the proportion of flare stars by color indices in the quadrant IV by dividing the stars into two groups: (1) main sequence stars and (2) subgaints and redgiants.
We defined the selection criteria based on the color-magnitude diagram (in Figure~\ref{fig:flaring_fraction_CMD}) and the methodology for distinguishing hot main sequence stars and subgaints outlined by \cite{2021MNRAS.506.5681D} in their Figure 7, as follows:
\begin{itemize}
    \item{Main Sequence Stars:}
    \begin{itemize}
        \item \textit{Hot Main Sequence Stars:} $4 \ge M_{G}$ and $-0.25 \le BP-RP \le 1$
        \item \textit{Cold Main Sequence Stars I:} $4 \le M_{G} \le 10$ and $BP-RP \ge 0.5$
        \item \textit{Cold Main Sequence Stars II:} $4 \ge M_{G}$ and $BP-RP \ge 1.5$
    \end{itemize}
    These criteria allow us to select nearly all main sequence stars while excluding most hot subdwarfs, giants, and white dwarfs.
    \item {Subgiants and Red giants:}
    \begin{itemize}
        \item \textit{} $M_{G} < 4$ and $BP-RP > 1$
    \end{itemize}
\end{itemize}}
{
Figure~\ref{fig:flare_stars_proportion_histogram} shows the distribution of star counts and the proportion of flare stars by color indices. We identified 16,923 flare stars out of 371,551 main sequence stars, indicating that approximately 4.6\% exhibit flare activity. The proportion of flare stars increases with redder color indices, peaking at $BP-RP$ values between 3 and 3.5, which correspond to spectral types M4 and M5. Beyond M5-M6, the proportion of flare stars decreases but then shows an increase at M7. Among giant stars, including subgiants and red giants, approximately 0.7\% (357 out of 50,347 stars) exhibit flaring activity, as identified by our machine-learning models.
}
    
Low-mass main sequence stars are categorized into M, K, G, and F types based on their $BP-RP$ color indices. 
We used the color index ranges for these spectral types from \cite{2020pase.conf..226B}, as depicted in their Figure 3.
In this figure, Barron et al. classified the $BP-RP$ color indices for various spectral types by cross-matching available data from the SIMBAD database.
Specifically, the color index ranges are as follows: approximately 5.3 to 1.95 for M type stars, 1.95 to 1 for K type stars, 1 to 0.63 for G type stars, and 0.63 to 0.36 for F type stars.
By using these categorizations, we assessed the flare activity across various spectral types of stars.
M~dwarfs have the highest amount of flares, totaling 150,803. 
This was followed by K dwarfs with 50,434 flares, and G dwarfs with 33,861. 
The number of flares detected in F dwarfs was significantly lower, only 2,564 flares have been detected.
For mid-mass dwarfs (i.e., A-type stars), we categorize them by a color index range of 0.36 to -0.126 and the $M_{G}$ range from 3 to 1.
Based on this, we identified 573 flares from the A~dwarfs.
For hotter O and B~dwarfs, they have the color less than -0.126 and $M_{G}$ brighter than 1.
Only 12 flares have been detected in the stars that satisfy these criteria for color and brightness.

\subsection{Flare Parameters and Uncertainty Estimates due to the Limited Cadence}
\label{subsec:flare_parameters_uncertainties}
The amplitudes and durations of flares detected in this work have been calculated before the models' flare identification task. 
We calculate the TESS response energy of all detected flares by using the equation:
\begin{equation}
E_{f} = ED \times L_{*},
\label{eq:flare_energy}
\end{equation}
where $E_{f}$ represents the flare energy, $ED$ denotes the equivalent duration of the flare in second, and $L_{*}$ is the star's quiescent luminosity under the TESS band transmission. 
The baseline flux at TESS magnitude of 0 was determined to be $4.03 \times 10^{-6}~\text{erg}~\text{s}^{-1}~\text{cm}^{-2}$ \citep{2015ApJ...809...77S}. 
Based on this, we utilize the TESS magnitude data and distance information from TIC~v8.2 \citep{2021arXiv210804778P} and Gaia~DR3 catalogs to estimate the quiescent TESS luminosity for each star by using the equation $L_{*}=F_{*}\times4\pi d^{2}$, where $F_{*}$ is the quiescent flux and $d$ is the distance of the star.
{Please note that flare energy estimated using this equation is the energy observed under the TESS bandpass, not the total energy or bolometric energy of the flare.}


{To evaluate the uncertainties in flare amplitude, duration, and energy resulting from limited observation cadence and varying signal-to-noise scenarios, we employed a simulation-based approach. 
At the beginning of the simulation, we generated noiseless standard flare profiles at a 0.1-second cadence for various durations, ranging from 14 minutes to 720 minutes, by utilizing the formulae outlined by \cite{2017SoPh..292...77G}, specifically their Equations 1 through 4. 
The input duration cases were as follows: [14, 30, 40, 60, 70, 90, 100, 120, 130, 140, 220, 290, 360, 430, 510, 580, 650, 720] minutes. 
For each input duration, we simulated light curves with a 2-minute cadence by applying a rolling average with a randomly shifted 2-minute box to the 0.1-second cadence standard flare profile. 
This process was repeated a thousand times to create a thousand 2-minute cadence light curve profiles for each input duration.
We injected Gaussian noise with varying standard deviation levels into every 2-minute cadence profile. 
The standard deviation levels used in our simulation were: [0.1, 0.05, 0.01, 0.005, 0.001, 0.0005]. 
Each standard deviation case was applied a thousand times as well. 
Therefore, there were 6,000 2-minute flare light curve profiles for each input duration case, and 108,000 profiles for all duration and standard deviation cases.
We then calculated the flare amplitudes, durations, and equivalent durations from these 2-minute cadence profiles and compared the results with the corresponding values derived from the noiseless 0.1-second cadence standard flare profiles.
}

The extent to the cadence impact on flare duration is influenced by both the actual duration of the flare and the noise level of the light curve.
For the short flares, the duration in 2-minute profile tends to be overestimated under any noise levels.
In contrast, the observed duration of these events in 2-minute profiles is likely to be  underestimated, and the situation becomes more severe as the noise level of light curve increases.
The flare amplitude from 2-minute profiles is typically significantly underestimated when the duration of a flare is short, which as expected. 

The impact of observation cadence on the derived flare energy, in terms of equivalent duration, is not as strong as it is on the other two parameters, with deviations generally within 5\% or less.
In most instances, if a flare has a long duration and the noise level of the 2-minute light curve the flare within is low, its estimated energy is typically a lower limit of the actual energy.
For shorter flares, the noise level of the light curve plays a greater rule than the observation cadence.
Under conditions of low noise, the estimated energy for such an event is likely to represent the upper limit, rather than the lower limit, of the actual energy.

\begin{figure}
    \centering
    \epsscale{1.}
    \plotone{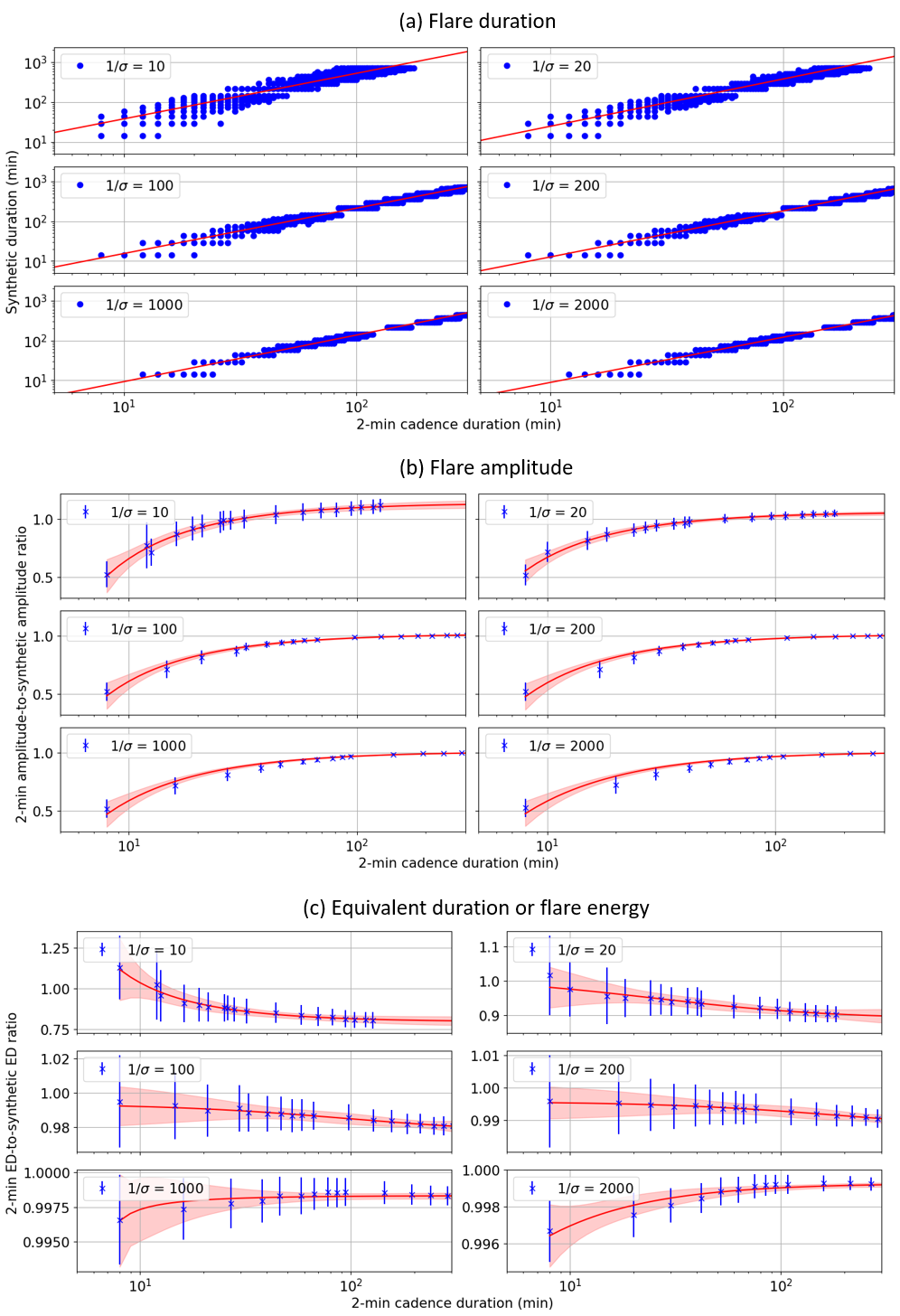}
    \caption{ (a) shows The relationship between the calibrated flare duration and observed duration (Eq.~\ref{eq:duration_calibration}). (b) and (c) illustrate the factors for calibrating flare amplitude and energy as the function of observed duration, as given by Eq.~\ref{eq:energy_amplitude_calibrate}. The coefficients for different light curve noise level are consolidated in Table~\ref{tab:formulae_coefficients_for_parameters_calibration}. 
    }
    \label{fig:flare_parameters_calibration}
\end{figure}

We found the interpretable relationship as a function of 2-minute observed duration in minute to calibrate these flare parameters under different light curve noise levels. 
The relationship between 2-minute observed duration and the actual duration of the synthetic profiles follows the power-laws (see Figure~\ref{fig:flare_parameters_calibration}-a),
\begin{equation}
\delta t_{calibrated} = b \times \delta t_{observed}^{a}
\label{eq:duration_calibration}
\end{equation}
and the coefficients, a and b, are listed in Table~\ref{tab:formulae_coefficients_for_parameters_calibration}.
For flare amplitude and energy (see Figure~\ref{fig:flare_parameters_calibration}-b and -c), the relationship between the calibration factor, which will be used to calibrate energy and amplitude, and the 2-minute observed duration can be expressed as
\begin{equation}
f = \frac{1}{a + b \times x} + c
\label{eq:energy_amplitude_calibrate}
\end{equation}
The coefficients, a, b, and c, of the formulae under different noise levels are also consolidated in Table~\ref{tab:formulae_coefficients_for_parameters_calibration}.
The calibrated flare amplitude and energy can be obtained with 
\begin{equation}
y_{cal} = \frac{y_{obs}}{f}
\label{eq:energy_amp_calibration}
\end{equation}
where $y_{obs}$ represent the observed amplitude and energy in the 2-minute TESS light curve, and $y_{cal}$ represents the calibrated values of these parameters.




\section{Discussion}
\label{sec:discussion}

Consequently, we introduce a catalog of flares identified through our machine learning models, alongside a corresponding catalog of flare stars. 
A sample from the flare catalog is illustrated in Table~\ref{tab:flares_catalog}, which lists observed parameters such as the timing of the flare peak in Barycentric TESS Julian Date (BTJD), duration, peak amplitude, flare energy, etc.
Similarly, Table~\ref{tab:flaring_stars_catalog} showcases a selection from the flare star catalog, detailing basic stellar parameters, brightness magnitude, rotation period, number of flares observed, among others. 

In this section, we will delve into the flare behavior across various star types by interpreting the flare frequency distributions considering potential observational biases which may arise due to limitations in the photometric precision and the data collection methods employed.

We will also discuss how the flare activities of different types of stars are correlated with their rotation periods. 
Highlighting the capabilities of our models, we will present intriguing cases and anomalies they have uncovered. 
Finally, we will conclude with insights into future research directions in anomaly detection using machine learning techniques.
{For broader accessibility and to facilitate further research, our models will be publicly available online at \dataset[Zenodo]{https://doi.org/10.5281/zenodo.13149385}.}

\subsection{Flare Energy, Amplitude, and Duration across different spectral types}
\begin{figure}
    \centering
    \epsscale{1.2}
    \plotone{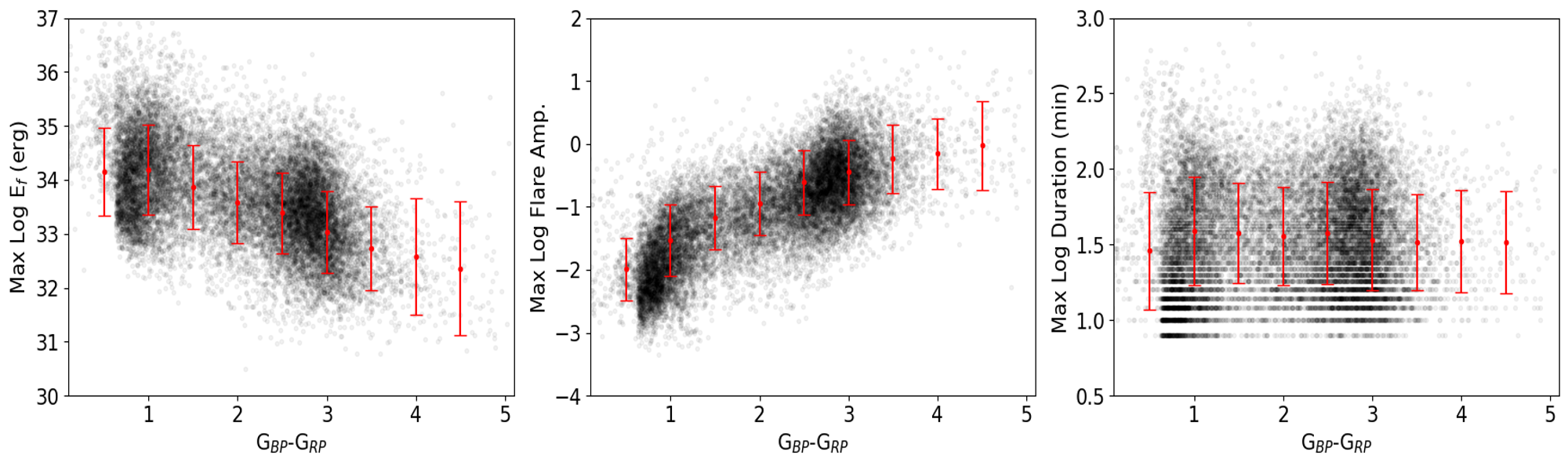}
    \caption{The relationships between these three flare parameters and the $BP-RP$ color indices. 
    The left panel illustrates that the maximum log flare energy tends to decrease with increasing $BP-RP$ color index.
    The middle panel demonstrates that the maximum log flare amplitude increases as the $BP-RP$ color index increases.
    The right panel shows that the maximum log flare duration does not display a strong trend with the $BP-RP$ color index. The red points with error bars represent the binned averages and their corresponding standard deviations, highlighting the overall trends and variances within the data.
    }
    \label{fig:flare-vs-colors}
\end{figure}

We estimate the maximum values of the calibrated energy, amplitude, and duration for each flaring star we have analyzed.
Figure~\ref{fig:flare-vs-colors} shows that the relationships between these three flare parameters and the $BP-RP$ color indices.
We observe that there are the relatively strong correlations in the energy--color and amplitude--color relationships.
Stars with redder colors, (i.e., cooler and lower massive stars) generally exhibit lower maximum flare energies compared to hotter stars.
On the contrary, the maximum flare amplitudes tend to be larger in cool stars and smaller in hot stars.
These findings are in agreement with the previous results of stellar flares analysis from the Kepler \citep[e.g.,][]{2019ApJS..241...29Y}, the K2 \citep{2019ApJ...873...97L}, and the Next Generation Transit Survey \citep{2021MNRAS.504.3246J} at different observation cadences.
In addition, we find the maximum flare duration does not display a strong trend with the $BP-RP$ color index, suggesting that this parameter of flare is relatively independent of the spectral types of stars.

\subsection{Flare frequency distribution with detection sensitivity calibration for stars from M- to A-types}
\label{sec:FFD}
The flare activity of stars can be characterized by its flare frequency distribution, which follows a linear relationship in logarithmic form:
\begin{equation}
\label{eq:CFD}
    \log N_{f} = a + \beta \log E_{f}, \label{eq:eq11}
\end{equation}
{
where \(N_{f}\) represents the cumulative frequency of flares with released energy \(E_{f}\).}
The flare frequency distribution can also be expressed in its differential form as follows:
\begin{equation}
    dN_{f} \propto E_{f}^{-\alpha} \, dE_{f}, \label{eq:eq12}
\end{equation}
{
where \(dN_{f}\) denotes the number of flares with energies in the range \(E_{f}\) to \(E_{f} + dE_{f}\), and \(\alpha = 1 - \beta\) \citep{2014ApJ...797..121H}.}
One explanation for why the stellar flare frequency as a function of flare energy follows a power--law distribution is the Self-organized Criticality (SOC) in the magnetic field lines near the stellar surface \citep[e.g.,][]{2021ApJ...910...41A}.
This theory briefly states that the stellar magnetic field reaches a critical state of complexity and instability, where even small perturbations can trigger magnetic reconnections that release energy in the form of flares with energy following a power-law distribution.
The power-law slope of \(\alpha = 2\) represents a critical value that determines whether the total energy of a size distribution diverges at the point of the largest flare (\(\alpha < 2\) or  \(\beta > -1\)), indicating a dominance of larger events, or at the smallest flare (\(\alpha > 2\) or  \(\beta < -1\)), where numerous smaller events, often referred to as nanoflares, significantly contribute to the energy distribution.

\begin{figure}
    \centering
    \epsscale{1.1}
    \plotone{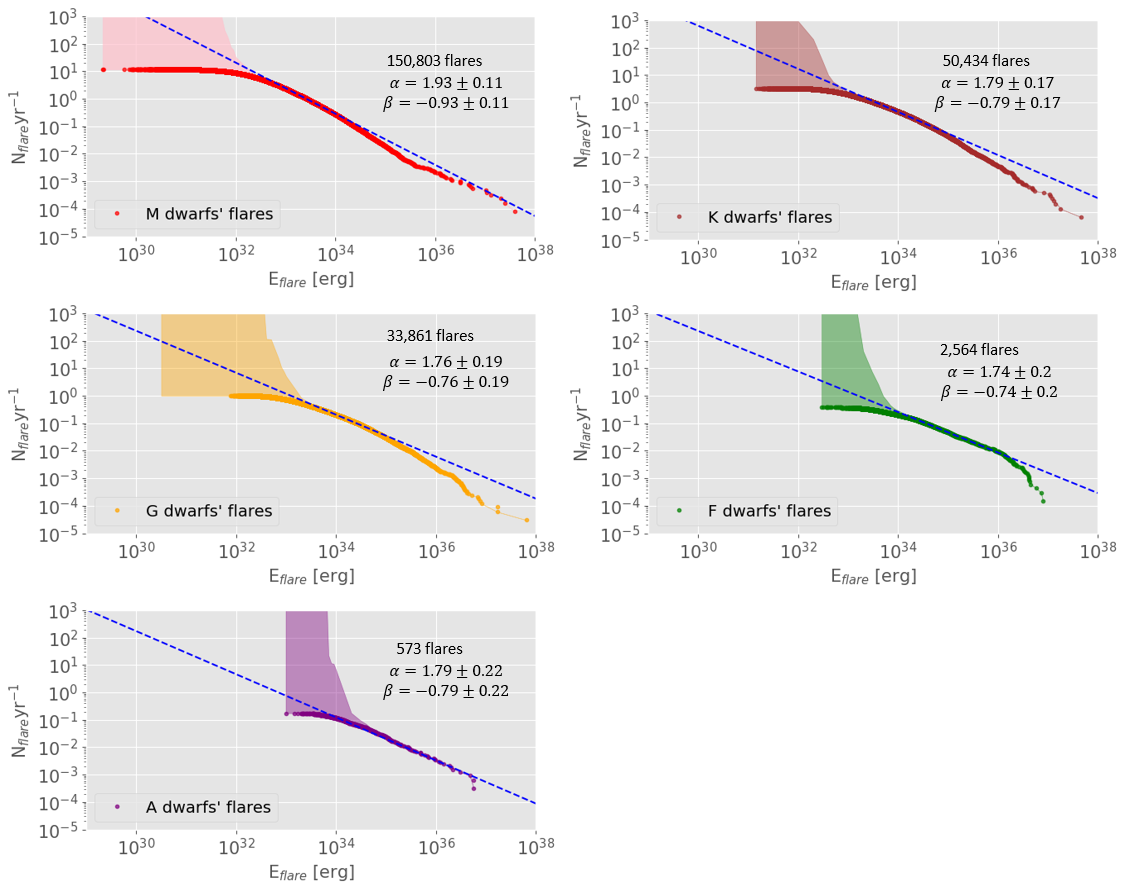}
    \caption{
    The cumulative flare frequency distributions of M, K, G, F, and A-type dwarf stars. 
    The shaded areas represent the deduced frequency uncertainties due to the detection incompleteness of small-amplitude events. The best-fit power-law indices ($\alpha$) are not only highlighted in texts with the blue dashed lines but listed in Table~\ref{tab:power-law-FFD}.
    }
    \label{fig:flare_FFD_MKGFA}
\end{figure}

Figure~\ref{fig:flare_FFD_MKGFA} shows the cumulative flare frequency distributions of the flares observed in M, K, G, F, and A-type main sequence stars.
Choosing an appropriate energy range to fit the power-law slope is challenging due to issues with detection completeness. 
The frequency distribution at lower energies is often incomplete, affected by limitations such as the sampling interval and the photometric precision of the survey. 
To address these uncertainties at lower energy levels, we conducted an injection and recovery test before calculating the power-law slope index.
We employed Equations 1 to 4 from \cite{2017SoPh..292...77G} once again to construct the synthetic flare profiles used in this task. 
The duration of each synthetic flare was fixed at about 14 minutes.
The baseline amplitude of 0.005 was calculated for a flare energy of $1\times10^{29}$~erg on a late-M~dwarf ($T_{eff}=~2500~K~and~R=0.1~R_{\odot}$) with the luminosity of $~1.4\times10^{29}~erg~s^{-1}$ under the TESS response.
The amplitude of the synthetic flares was made proportional to the input energy, e.g., the flare with the energy of $1\times10^{30}$~erg has the synthetic amplitude of 0.05 if it is observed in a late-M~dwarf.
Furthermore, the baseline amplitude for a $1 \times 10^{29}$~erg flare varies across different stellar types. 
For instance, if such a flare occurs on an M0 dwarf, which has a luminosity 825 times greater than that of a late-M dwarf (i.e., $1.2 \times 10^{32}$~erg~s$^{-1}$), the resulting amplitude would be approximately $6 \times 10^{-6}$.
For simplicity, we conducted tests only for the '0' subtype across every spectral type of star, and the luminosities of these stars under the TESS response are as follows: $1.2 \times 10^{32}$~erg~s$^{-1}$ for M0, $4.9 \times 10^{32}$~erg~s$^{-1}$ for K0, $1.4 \times 10^{33}$~erg~s$^{-1}$ for G0, $6.0 \times 10^{33}$~erg~s$^{-1}$ for F0, and $2.2 \times 10^{34}$~erg~s$^{-1}$ for A0.

The input flare energy range is defined as $\log E_f = [29, 38]$ with a bin size of 0.1. 
The synthetic flare profiles are generated with a cadence of 0.1 seconds. 
We generated the 2-minute cadence flare model profiles using the methodology outlined in Section~\ref{subsec:flare_parameters_uncertainties}. 
A consistent {Gaussian noise with a standard deviation} of 0.0005, reflective of the general noise observed in the actual TESS data, was injected into all 2-minute flare model profiles.
We generated 1,000 such profiles for each input energy level on each type of star.
These were then passed to our machine-learning models to carry out the recovery rates assessment.
The recovery rates for each case were calculated as the ratio of number of flare models identified as 'True' to the 1,000.
These recovery rates were subsequently used to calibrate the cumulative flare frequencies for all types of stars.
This has been done by dividing the observed flare frequencies by the recovery rates.
We considered these calibrated values as upper limits due to the simplistic assumptions in our recovery task setup. 
The uncertainties were determined by the difference between these calibrated flare frequencies and the observed flare frequencies, as depicted in the shaded areas in Figure~\ref{fig:flare_FFD_MKGFA}.

We derived the power-law slopes $\alpha$ from the $\beta$ slopes estimated from the least-squares fit of the flare frequency distributions with the Eq.~\ref{eq:CFD}. 
The resulting power-law slopes $\alpha$ for the flare frequency distributions across different spectral types are as follows: $1.93\pm0.11$ for M~dwarfs, $1.79\pm0.17$ for K~dwarfs, $1.76\pm0.19$ for G~dwarfs, $1.74\pm0.2$ for F~dwarfs, and $1.79\pm0.22$ for A~dwarfs (see Table~\ref{tab:power-law-FFD}).
We see that the power-law slope gradually decreases from M to F dwarfs, indicating a less steep distribution, and then increases for A dwarf flares.
{However, when considering uncertainties, the power-law slopes for A through K dwarfs are consistent with each other.
Compared to what \cite{2019ApJ...873...97L} observed for M to G-type flare stars from the K2 long-cadence (30~min) data, which are $1.82\pm0.02$ for M~dwarfs, $1.86\pm0.02$ for K~dwarfs, and $2.01\pm0.003$ for G~dwarfs, our results for M and K dwarfs are consistent with their results except for G~dwarfs.
The relatively large difference for G~dwarfs may be because they did not include uncertainties from missing small flares.}
Also, the cadence of K2 data is 15 times worse than the TESS 2-minute cadence, suggesting that the incompleteness issue could be much worse as well.
On the other hand, \cite{2023A+A...669A..15Y}, from their stellar flare statistics in the first two years of the TESS survey, observed a phenomenon in flares of single star from M-type to G-type that is comparable to our results in this study.
The power-law $\alpha$ slopes of flare frequency distributions they derived for M, K, and G-type single stars are $1.88\pm0.16$, $1.81\pm0.17$-, and $1.77\pm0.4$, respectively.
{For the single stars, our measurements for G through M~dwarfs are in better agreement with theirs than with those of \cite{2019ApJ...873...97L}.}

From the Kepler's long-cadence survey, \cite{2019ApJS..241...29Y} reported a power-law slope $\alpha$ of \(2.11 \pm 0.09\) for flares on F-type stars.
Similarly, from a larger sample size, \cite{2023A+A...669A..15Y} obtained the power-law slope $\alpha$ of \(2.28 \pm 0.20\) for all non-EBs, F-type stars.
These values are significantly higher than our result of \(1.74 \pm 0.20\). 
This discrepancy may arise from their calculations not accounting for uncertainties due to the incompleteness of small flares. 
By fitting the cumulative flare frequency distribution starting from \( \log(E_f) = 35.6 \), similar to the initial energy used by \cite[][their Figure~19]{2023A+A...669A..15Y}, and excluding smaller flares, we obtained an \(\alpha\) of \(2.23 \pm 0.24\), which aligns closely with their results.

For A-dwarfs' flares, \cite{2019ApJS..241...29Y} reported a power-law slope \(\alpha\) of \(1.12 \pm 0.08\). 
Similarly, \cite{2020ApJ...905..110B} determined the \(\alpha\) for A-type star flares to be \(1.26 \pm 0.56\).
Later, \cite{2023A+A...669A..15Y} obtained the $\alpha=1.76\pm0.19$ for all A-type stars and $\alpha=1.86\pm0.15$ for non-EBs A-type stars.
With consideration of the frequency uncertainties in the small flare regime, we derived the power-law slope $\alpha$ of $1.79\pm0.22$ from an even larger dataset in this work for A~dwarf flares.
When excluding the small flares, the slope $\alpha$ becomes $1.86\pm0.25$.
As a result, our power-law slope measurement for the A~dwarf flares is not influenced by the uncertainties of detection incompleteness as severely as F~dwarf flares, thus further confirming the findings of \cite{2023A+A...669A..15Y}.

\subsection{Stellar Flare Activity and Rotation Period} \label{subsec:flare_activity_and_rotation_period}
It's been well known that there is a close relationship between a low-mass star's rotation and its magnetic activity such as flares \citep{2017ApJ...834...92C, 2018ApJ...867...78C, 2019ApJ...873...97L, 2019ApJ...871..241D}
The flare energy of a star is strongly correlated to its rotation period, that is, stars with shorter rotation period tend to generate higher-energy flares than those with longer periods.
In addition, stellar magnetic activity, often linked to magnetic fields in the corona, chromosphere, and photosphere, results in the emission at high-energy frequencies (e.g., X-ray) \citep{2003A&A...397..147P, 2011ApJ...743...48W, 2019A&A...628A..41P}.
The X-ray luminosity of a star is also following a trend similar to that seen in the flare energy--rotation correlation. 
However, when the rotation speed reaches a certain threshold, the X-ray luminosity stops increasing, entering a saturation regime.
The saturation regime is believed to be a manifestation of magnetic saturation \citep{2009ApJ...692..538R}. 

\begin{figure}
    \centering
    \epsscale{0.95}
    \plotone{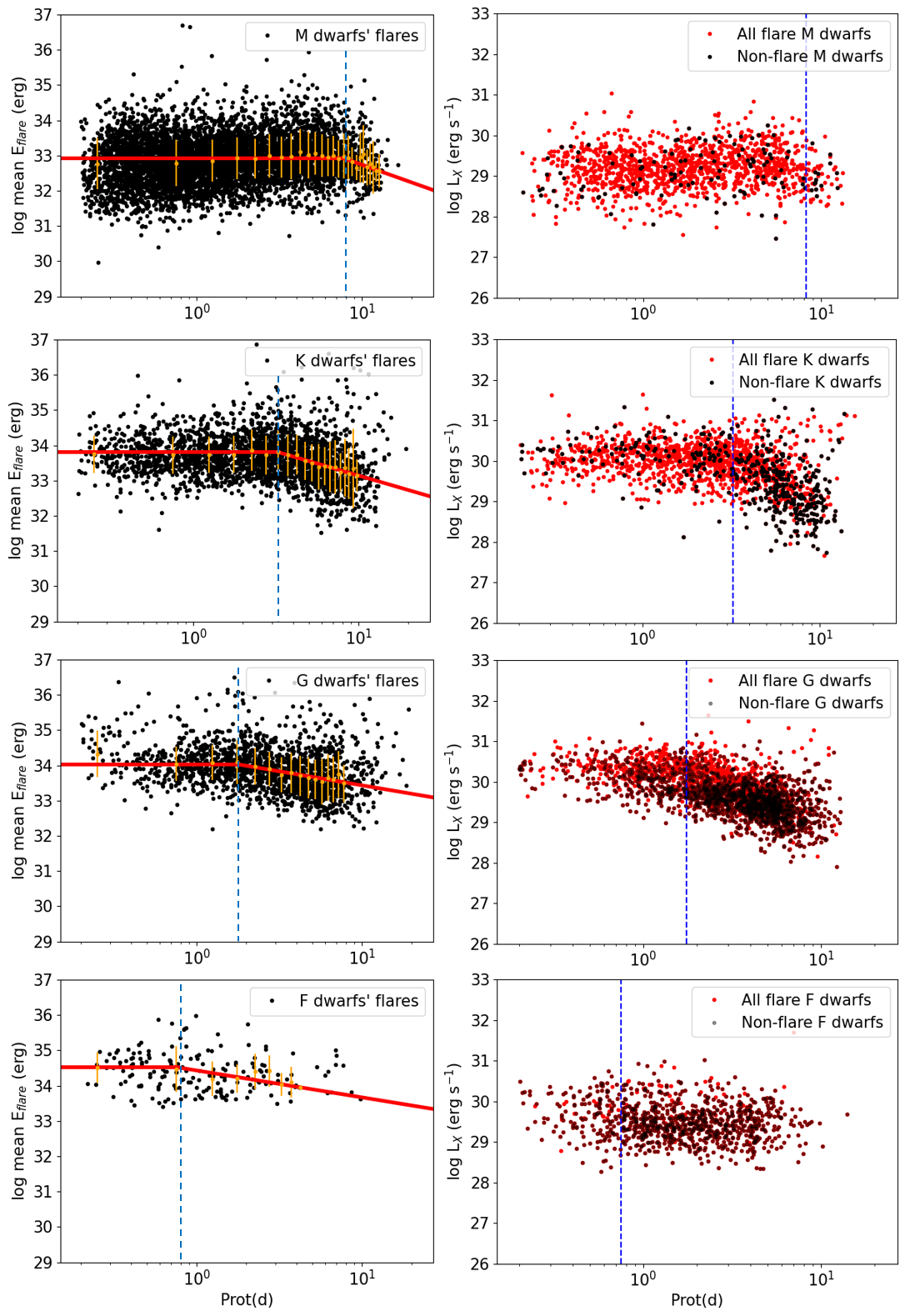}
    \caption{
    The panels on the left column shows the relationships between the logarithmic mean flare energy and rotation period for M, K, G, and F-type dwarf stars.  The panels on the right represent the logarithmic X-ray luminosity--rotation period relationships.
    The X-ray luminosity of our stellar samples are estimated from the data provided by \cite{2022A&A...664A.105F}. 
    The red and black dots represent the stars with and without flare detection in this study, respectively.
    The vertical blue dashed lines pinpoint the transition turnoff values in term of rotation period that separated the stars into saturated and unsaturated regimes. 
    The red solid lines on the left panels represent the best-fit power-law slope of flare energy--rotation period relationships in the unsaturated regimes. 
    These numbers of transition turnoff periods and power-law slopes can be found in Sec~\ref{subsec:flare_activity_and_rotation_period} and Table~\ref{tab:flare_vs_rotation}.
    }
    \label{fig:flare_eng_vs_Prot_MKGF}
\end{figure}

In this work, we have calculated the rotation periods for 120,782 stars, all of which meet the validity criteria outlined in Section \ref{sec:rotation_period}. 
We estimated the mean values of all detected flares for these stars and examined the correlation between mean flare energy and rotation period for M, K, G, and F-type main-sequence stars (Figure~\ref{fig:flare_eng_vs_Prot_MKGF}).
There are clear trends between flare energy and rotation period, and the saturation regimes for different spectral types of stars can be easily identified. 
The unsaturated flare energy–rotation period relationships follow a power-law.
The power-law slopes for M, K, G, and F-type dwarfs are estimated to be $-1.67\pm0.21$, $-1.37\pm0.1$, $-0.8\pm0.11$, and $-0.76\pm0.27$, respectively (see Table~\ref{tab:flare_vs_rotation}).
The slopes of these power-law relationships become less negative from cooler to hotter stars, indicating that magnetic activity decreases less rapidly with slower rotation periods in hotter stars compared to cooler stars.

The transition turnoff rotation periods that separates the saturated and power-law regimes for these types of stars are also determined: $P_{rot}=8\pm0.25$~days for M dwarfs, $P_{rot}=3\pm0.25$~days for K dwarfs, $P_{rot}=2\pm0.25$~days for G dwarfs, and $P_{rot}=0.75\pm0.25$~days for F dwarfs; the uncertainties are based on the bin width of the fitted data points (Table~\ref{tab:flare_vs_rotation}).
\cite{2019ApJS..241...29Y} and \cite{2019ApJ...873...97L} observed and obtained the turnoff rotation periods for M~dwarfs of $~10$~days from the Kepler/K2 long-cadence observation.
While \cite{2019ApJ...873...97L} failed to determine the turnoff point for K-dwarfs, \cite{2019ApJS..241...29Y} found a value of $~5$~days.
From the X-ray activity versus rotation period relationship, \cite{2019A&A...628A..41P} identified turnoff periods of approximately 3 days for K dwarfs and 2 days for G dwarfs, which are in good agreement with our measurements.
With a larger sample of fast-rotating stars, we determined a shorter turnoff period for F~dwarfs than the approximate 2~days derived by \cite{2019A&A...628A..41P}.
The transition turnoff rotation periods is longer for cooler stars and shorter for hotter stars. 
This may suggest that cooler stars can maintain high levels of magnetic activity over a broader range of rotation periods compared to hotter stars.

The column on the right in Figure~\ref{fig:flare_eng_vs_Prot_MKGF} shows the relationship between the X-ray luminosity and the rotation periods for the stars in the corresponding spectral types.
The X-ray luminosity of our stars was estimated using distance information provided from \citep{2021arXiv210804778P} and X-ray flux data in the 0.1 keV to 2.4 keV range (5.2 to 124~$\AA$ ) from the ROSAT observatory \citep{2022A&A...664A.105F}.
The X-ray--rotation correlations are nearly identical to those seen in the flare energy--rotation relationships, which further confirms that there is a strong connection between flare and X-ray emission (i.e., coronal heating) associated with the magnetic dynamos in low-mass stars.
However, we notice that a considerable fraction of stars emitting X-rays have no detected flares in the TESS light curves, and this fraction increases for hotter stars: 9.3\% of M dwarfs, 28\% of K dwarfs, 69\% of G dwarfs, and 94\% of F dwarfs emit X-rays without any flare detection in this study.
The discrepancy could be partly due to TESS detection limits for flares. 
X-rays may reveal a wide range of flare activities, some of which may fall below the TESS detection threshold in its visible band.
This limitation is particularly significant for hotter stars like G and F~dwarfs, which tend to have lower optical flare amplitudes relative to their quiescent brightness.
This has been addressed in depth in the previous Section (Sec~\ref{sec:FFD}).
In short, stars without detectable flares in TESS light curves could still have high X-ray activity due to continuous coronal heating or unresolved small-scale, transient magnetic activity.
The increase in X-ray-emitting stars without detected flares for hotter stars may also reflect the evolutionary differences in magnetic activity, which requires further investigation in future studies.

\subsection{Relationship between Flare activity and X-ray luminosity}
We further examined the relationship between the flare activity and the X-ray luminosity for M, K, G, and F-type stars.
To this end, we first calculated the mean flare luminosity by using the the following equation
\begin{equation}
\label{eq:mean_L_f}
\overline{L_f} = \frac{1}{N} \sum_{i=1}^{N} \frac{E_{fi}}{\Delta t_i},
\end{equation}
where \( \overline{L_f} \) is the mean flare luminosity in ergs~s$^{-1}$, \( N \) is the total number of detected flares for a star, \( E_{fi} \) represents the energy of the \(i\)-th flare in ergs, and \( \Delta t_i \) is the calibrated duration of the \(i\)-th flare in seconds.
We note that all the luminosity and energy are under the TESS instrument response and are not in bolometric units.

\begin{figure}
    \centering
    \epsscale{0.85}
    \plotone{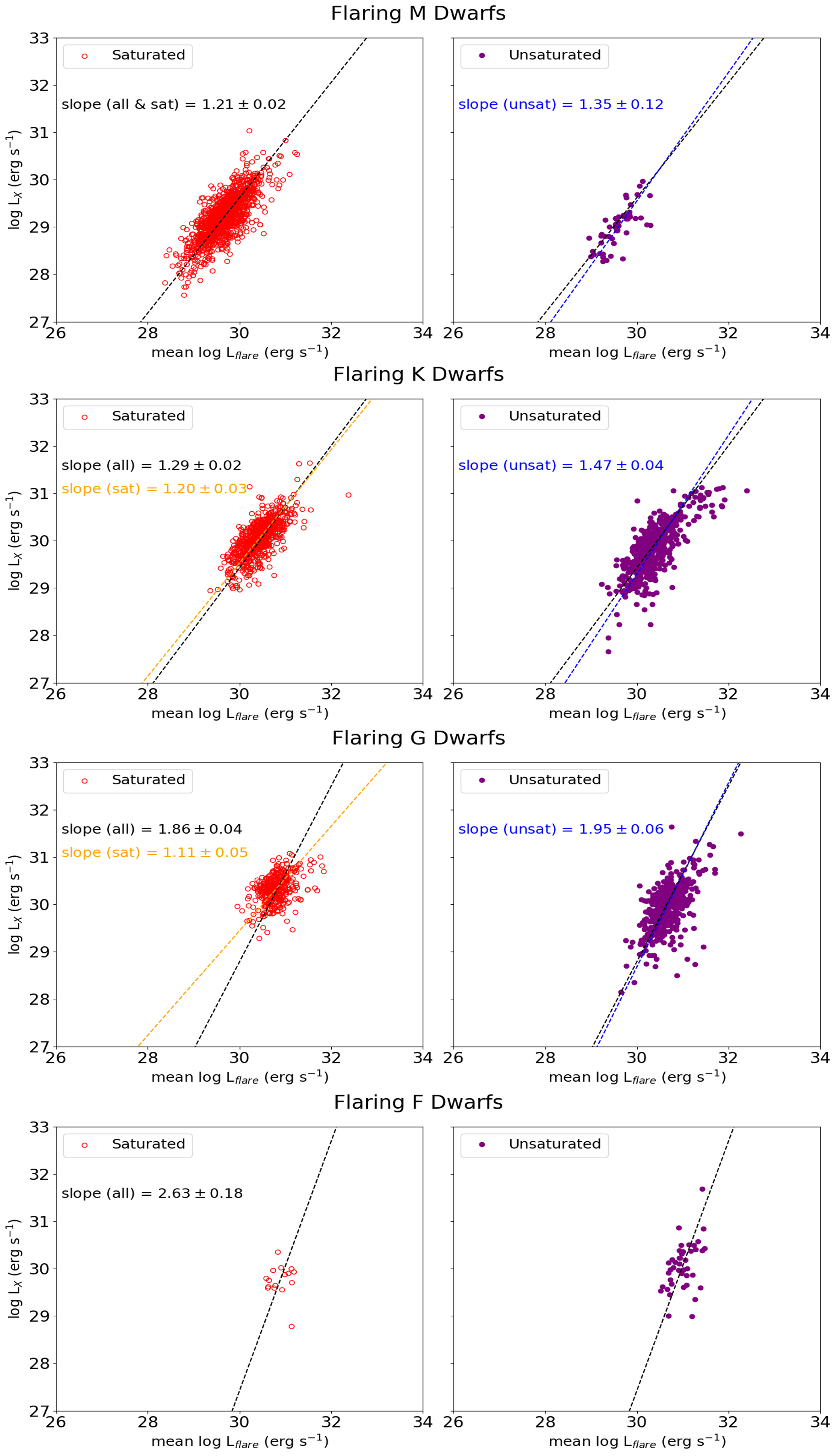}
    \caption{
    The log-log relationships between X-ray luminosity and mean flare luminosity for M, K, G, and F-type dwarf stars.
    The red unfilled circles represent the stars in the saturated regimes, and the purple circles are the stars in the saturated regimes.
    The black dashed lines represent the best-fit slopes for stars in both saturated and unsaturated regimes. 
    The yellow and blue dashed lines are the best-fit slopes for saturated and unsaturated stars, respectively. 
    }
    \label{fig:Lx-vs-Lf-MKGF}
\end{figure}

Figure~\ref{fig:Lx-vs-Lf-MKGF} shows that there are positive correlations between mean flare luminosity and X-ray luminosity in logarithmic space for flaring M, K, G, and F-type dwarfs.
We determined the best-fit linear functions to interpret these correlations as follows
\begin{equation}
\label{eq:flare-Xray-correlation_M}
\log(L_X) = (-6.86 \pm 0.03) + (1.21 \pm 0.02) \log(\overline{L_f})~\text{for M~dwarfs,}
\end{equation}
\begin{equation}
\label{eq:flare-Xray-correlation_K}
\log(L_X) = (-9.31 \pm 0.03) + (1.29 \pm 0.02) \log(\overline{L_f})~\text{for K~dwarfs,}
\end{equation}
\begin{equation}
\label{eq:flare-Xray-correlation_G}
\log(L_X) = (-26.98 \pm 0.06) + (1.86 \pm 0.04) \log(\overline{L_f})~\text{for G~dwarfs, and}
\end{equation}
\begin{equation}
\label{eq:flare-Xray-correlation_F}
\log(L_X) = (-51.34 \pm 0.27) + (2.63 \pm 0.18) \log(\overline{L_f})~\text{for F~dwarfs.}
\end{equation}
For earlier-type stars, a steeper slope indicates that X-ray luminosity increases more rapidly with an increase in flare luminosity compared to later-type stars. 
One explanation for this behavior could be that a greater proportion of earlier-type stars are in the unsaturated regime compared to later types. 
Within this unsaturated regime, the steeper correlation likely results from more efficient atmospheric and coronal heating processes.
\cite{2022ApJ...927..179T} reported the power-law slopes of the correlation between X-ray irradiance and magnetic flux observed in the Sun across different solar activity cycles. 
They found that the power-law slope reaches a minimum of \(0.90 \pm 0.05\) during the solar maximum period from September 2012 to July 2015, compared to a slope of \(1.18 \pm 0.11\) during the solar minimum period from December 2017 to February 2020. 
They proposed that this could be due to the partial saturation of atmospheric heating during solar maximum.

For M~dwarfs, the slope of the \( \log(L_X) - \log(\overline{L_f}) \) correlation in the saturated regime is \(1.21\pm0.02\), which is the same as the value we determined for all flaring M dwarfs. 
This suggests that saturated stars dominate the overall sample, as expected. 
In contrast, the slope for stars in the unsaturated regime is \(1.35\pm0.13\), which is steeper than that observed for saturated stars.
{For K dwarfs, the slopes within the saturated and unsaturated regimes are \(1.20 \pm 0.03\) and \(1.47 \pm 0.04\), respectively.}
{The difference in slopes between saturated and unsaturated regimes for K dwarfs becomes larger than that of M dwarfs.}
For G~dwarfs, the slope in the saturated regime is \(1.11\pm0.05\), and the slope in the unsaturated regime is \(1.95\pm0.06\), {exhibiting an even larger difference than that of K dwarfs.}
We don't estimate the slope in the saturated regime for F~dwarfs because of the small sample size (only 16 saturated F~dwarfs have reported X-ray luminosities). 
However, the slope in the unsaturated regime remains the same as the value we have obtained for all F~dwarfs.

For all spectral types, the slopes in the unsaturated regime are steeper than those in the saturated regime.
This may confirm our previous explanation for the slope variations, suggesting that coronal heating efficiency by flares is higher in the unsaturated regime. 
However, the differences in unsaturated slopes across spectral types imply intrinsic variations in magnetic dynamo processes or coronal heating mechanisms among different types of stars.


\subsection{Flare detection in other types of stars}





We have detected flares in stars beyond main-sequence stars with outer convective zones, including white dwarfs (WDs) and hot subdwarfs (sdO/sdB). 
Observations have shown that some white dwarfs display modulations in their light curves, suggesting the presence of spots and strong surface magnetic fields capable of triggering flares \citep[e.g.,][]{2017MNRAS.468.1946H,2021ApJ...914L...3H}.
In contrast, there is no observational evidence of similar spot-like modulations in the light curves of single sdO/sdB stars yet, and only a few of the known hot subdwarfs have detectable magnetic fields \citep{2022MNRAS.515.2496P}.

\cite{2024ApJS..271...57X} carried out a search for flares  in the WDs and sdO/sdB in the TESS 2-minute observation from Sector~1--69.
They found 834 flares in 135 white dwarfs and 182 flares in 58 sdO/sdB and further concluded that most of these events likely originated from nearby sources or companion stars.
We applied the selection criteria of $M_G > 7$ and $BP-RP < 1.5$ to identify WDs in the color-magnitude diagram, and we found that our models detected 620 flares in 72 WDs.
There are 15 of 72 flaring WDs in our samples that have been previously identified by \cite{2024ApJS..271...57X}.
For sdO/sdB stars, the selection criteria of $2 < M_G < 7$ and $BP-RP < 0$ is applied to identify these types of stars in our samples. 
Our models only detected 12 flares in four sdO/sdB stars, and one of them has been reported by \cite{2024ApJS..271...57X}.
The primary reason our flare detections are fewer than those reported by \cite{2024ApJS..271...57X} is that many of the flaring sources identified by Xing et al. are classified as non-quadrant IV stars in our V$_{amp}$-P$_{rot}$ (see Section~\ref{sec:rotation_period}) diagram and thus were excluded from our flare classification task. 
Additionally, known eclipsing binaries (EBs), which Xing et al. included, were also omitted from our analysis.

\begin{figure}
    \centering
    \epsscale{1.}
    \plotone{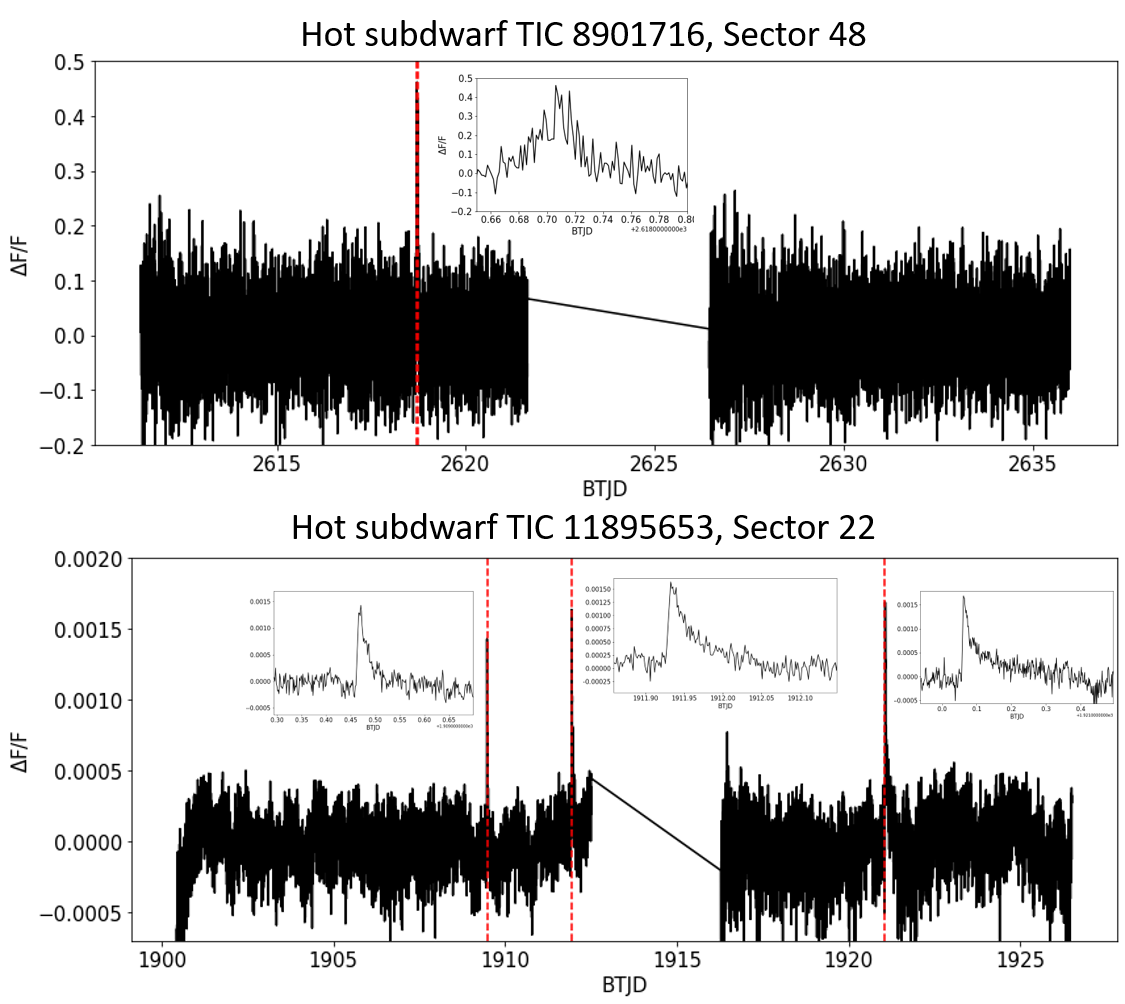}
    \caption{
    The flares we detected in two hot subdwarfs, TIC~8901716 in Sector~48 (top panel) and TIC~11895653 in Sector~22 (bottom panel). 
    }
    \label{fig:hot_subdwarf_flares}
\end{figure}

{Given the definition of potential contaminated sources based on the contamination ratio $R_{\text{cont}}$ we have introduced in Sec~\ref{subsec:flare candidates identification}, we observed that 31 of the 72 flaring WDs have a contamination ratio $R_{\text{cont}} > 0.1$.}
In addition, 59 of these 72 flaring WDs exhibit $BP-RP > 1$, suggesting the likely presence of a low-mass companion. 
There are only three likely single WDs with a $BP-RP < 1$ in our samples, and all of them have $R_{\text{cont}} > 0.2$, indicating a high probability of flux contamination from nearby sources.
For the flaring hot subdwarfs, two out of four have $R_{\text{cont}} > 0.1$, including TIC~157332788, previously reported by \cite{2024ApJS..271...57X}. TIC~8901716, not impacted by nearby sources, displayed a significant flare-like event in Sector~48 (top panel in Figure~\ref{fig:hot_subdwarf_flares}), with a high amplitude of about 0.35, lasting 30 minutes, and releasing an energy of $1.8 \times 10^{35}$~erg. 
The other uncontaminated hot subdwarf, TIC~11895653, proved to be the most flare-active in our study, exhibiting five flares. 
During observations in Sector~22, TIC~11895653 produced three flares in quick succession (bottom panel in Figure~\ref{fig:hot_subdwarf_flares}), each with similar amplitudes and energies of about 0.0014 and $4 \times 10^{32}$ erg, respectively, and lasting 34 minutes.
The origin of these flares in hot subdwarfs remains uncertain. 
A plausible hypothesis could be unresolved close-in low-mass stellar companion. 
Supporting this hypothesis, we found that TIC~11895653 is an identified X-ray source with a luminosity of $L_X = 3.5 \times 10^{28}$~erg/s \citep{2022A&A...664A.105F}, which hints at the potential activity of a companion.

\section{Conclusions}
\label{sec:conclusion}

In this study, we present a flare identification project using machine learning models on data from the TESS 2-minute survey across Sectors 1--72, resulting in catalogs of flare stars and stellar flares.
These models were trained using three different algorithms: Deep Neural Network (DNN), Random Forest Classifier (RF), and XGBoost Classifier. 
We defined four characteristics of flare light curve profiles as input features for the models to learn from. 
After training, we evaluated the performance of our models using four classic metrics---accuracy, precision, recall, and F1-score---and also conducted a flare recovery validation using the M dwarf flare observations reported by \cite{2020AJ....159...60G}. 
Our models successfully recalled more than 92\% of the flares detected by Gunther and identified approximately 2,000 additional small events that were previously undetected. 
This demonstrates that our models are not only functional but also more sensitive to detecting smaller flare events than the current state of the art.

More than 18,000 flare stars and nearly 250,000 flares have been detected by our models and are summarized in the catalogs along with their stellar and flare parameters. 
We extend and expand the previous studies of flares in the TESS database, mainly because of the higher sensitivity of our ML models to low-amplitude flare events.

We also reported the rotation periods for more than 120,000 stars, determined using the Generalized Lomb-Scargle periodogram and the Autocorrelation Function (ACF), as implemented in the astropy.timeseries package. 
Valid rotation periods were identified by the highest peak in the periodogram with a false alarm probability below 1\%, further confirmed by an R-square value exceeding 0.7 after fitting a 5th-order Fourier series to the phase curve. 
Consistency within 5\% between the Lomb-Scargle and ACF methods was required for a period to be accepted.

Due to the limitations of our light curve detrending approach, our analysis excluded known eclipsing binaries (EBs) and stars not classified as quadrant IV, which are characterized by light curve modulation amplitudes less than 0.2 and rotation periods over 0.2 days. EBs and stars from other quadrants will be thoroughly investigated in future studies using a more specialized detrending method.

We observe strong correlations in energy--color and amplitude--color relationships, with stars of redder colors (cooler and lower mass stars) generally exhibiting lower maximum flare energies compared to hotter stars. 
Conversely, cooler stars tend to have larger maximum flare amplitudes, while hotter stars have smaller amplitudes. 
These findings are consistent with previous studies from Kepler, K2, and the Next Generation Transit Survey. 
Additionally, we find that the maximum flare duration does not show a strong trend with the $BP-RP$ color index, suggesting that this parameter is relatively independent of the spectral types of stars.

We analyzed flare frequency distributions for M- to A-type stars using a combination of actual and synthetic flare data processed through advanced machine learning techniques. 
This approach enabled us to address the challenges associated with flare detection incompleteness at lower energy ranges, thus refining our understanding of the uncertainties in flare frequencies across different spectral types. 
Based on that, we derived power-law slopes \(\alpha\) that characterize flare behaviors for the stars: \(1.93 \pm 0.11\) for M dwarfs, \(1.79 \pm 0.17\) for K dwarfs, \(1.76 \pm 0.19\) for G dwarfs, \(1.74 \pm 0.20\) for F dwarfs, and \(1.79 \pm 0.22\) for A dwarfs (see Table~\ref{tab:power-law-FFD}). 
Notably, the reduced \(\alpha\) value for F-type dwarfs compared to previous studies (e.g., \(2.11 \pm 0.09\) by \cite{2019ApJS..241...29Y}) underscores the significance of accounting for detection incompleteness in flare energy calculations. 
Moreover, the alignment of our A-type star results with recent findings, such as those by \cite{2023A+A...669A..15Y}, confirms the robustness of our methodologies and the reliability of our conclusions.

We examined the connection between rotation periods and magnetic activity across M to F-type main-sequence stars and identified transition turnoff periods marking the shift to the saturation regmine as \(8 \pm 0.25\) days for M dwarfs, \(3 \pm 0.25\) days for K dwarfs, \(2 \pm 0.25\) days for G dwarfs, and \(0.75 \pm 0.25\) days for F dwarfs, i.e., the turnoff rotation periods decrease with earlier spectral types. 
Following these saturation regimes, the non-saturated flare energy-rotation relationships for these stars demonstrate power-law behavior with slopes of \( -1.67 \pm 0.21\) for M dwarfs, \( -1.37 \pm 0.1\) for K dwarfs, \( -0.8 \pm 0.11\) for G dwarfs, and \( -0.76 \pm 0.27\) for F dwarfs (see Table~\ref{tab:flare_vs_rotation}). 
These findings parallel the trends observed in X-ray luminosity versus rotation period, where similar saturation and power-law behaviors were identified, confirming that both flare energies and X-ray emissions are closely linked through stellar magnetic processes. 
Moreover, we observed a significant fraction of stars emitting X-rays without corresponding flare detections in TESS light curves, a phenomenon more apparent in hotter stars: 9.3\% of M dwarfs, 28\% of K dwarfs, 69\% of G dwarfs, and 94\% of F dwarfs. 
This discrepancy is likely influenced by TESS's detection limits for flares, as X-ray luminosity indicates the presence of nanoflares that speculatively leads coronal heating, most of which are too subtle to be detected by TESS's capabilities.
It may also suggest that potential evolutionary differences in magnetic activity, meriting further exploration in future studies.

We investigate the relationship between flare activity and X-ray luminosity in M, K, G, and F-type stars by calculating the mean flare luminosity, \(\overline{L_f}\), for each star and comparing it to the X-ray luminosity reported in the literature.
Our findings show a positive correlation in logarithmic space between mean flare luminosity and X-ray luminosity, with the best-fit linear functions revealing that earlier-type stars (F and G) have a steeper slope compared to later-type stars (M and K). 
This suggests that X-ray luminosity increases more rapidly with flare luminosity in earlier-type stars, likely due to a greater proportion of these stars being in the unsaturated regime where coronal heating is more efficient. 
Additionally, we observe that the slopes in the unsaturated regime are steeper than those in the saturated regime across all spectral types, further supporting the hypothesis of higher coronal heating efficiency by flares in the unsaturated regime. 
The variations in the slopes of unsaturated stars across different spectral types suggest intrinsic differences in magnetic dynamo processes or coronal heating mechanisms among stars of different types. 
Observationally, it may indicate the relative importance of 'quiescent' coronal heating and 'transient' coronal heating due to flare activity.


We have detected flares in stars beyond main-sequence stars with outer convective zones, specifically in white dwarfs (WDs) and hot subdwarfs (sdO/sdB). 
We identified 618 flares in 72 WDs and 12 flares in 4 sdO/sdB stars. 
To address potential contamination from nearby sources, we utilized the contamination ratio ($R_{\text{cont}}$) and flagged stars with $R_{\text{cont}} > 0.1$ as their data are likely influenced by the nearby sources. 
In our sample, 31 of 72 flaring WDs had $R_{\text{cont}} > 0.1$, and 59 exhibited $BP-RP > 1$, suggesting the presence of low-mass companions. 
For the flaring hot subdwarfs, two out of four had $R_{\text{cont}} > 0.1$.
The origin of these flares remains uncertain, but our findings suggest that unresolved close-in low-mass stellar companions might play a significant role.
Notably, TIC~11895653, the most flare-active hot subdwarf displaying five flares in this study, is an identified X-ray source, which hints at potential companion activity.

Finally, we conclude that our multi-algorithms machine learning approach is the highly effective tool for flare identification. 
Theoretically, the models we have trained should be applicable to light curve data observed at various cadences. 
Moving forward, we plan to explore the potential and scalability of our models by applying them to a range of data products, such as TESS full-frame image 30-minute data, TESS fast-cadence 20-second data, and time series data from upcoming surveys including PLATO \citep{2014ExA....38..249R}, ROMAN \citep{2015arXiv150303757S}, ET-2.0 \citep{2022arXiv220606693G}, and TAOS II \citep{2021PASP..133c4503H}. 
These efforts will enable us to enhance our understanding of stellar flare activity across different types of stars and observational platforms, broadening the applicability and impact of our models in the field of stellar astrophysics.
Moreover, we recognize that machine learning models have the potential to identify not only flares but also other transient events and objects in light curve data or similar time-series data. 
The specialized model could be used to detect stellar companions or exoplanets, unusual brightness dimming events like those observed in Tabby's Star \citep[e.g.,][]{2016MNRAS.457.3988B}, and outbursts due to mass accretion events in classical T Tauri stars \citep[e.g.,][]{2023ASPC..534..355F, 2023AJ....166...82L}.
In conclusion, machine learning offers a versatile and scalable solution for analyzing diverse datasets and revealing a wide range of astronomical phenomena.



\begin{acknowledgments}
This research is supported in part by grant No.~112-2917-I-008-001 from National Science and Technology Council (NSTC) of Taiwan and by grant No.~80NSSC21K0593 from National Aeronautics and Space Administration (NASA) of the US for the program "Alien Earths."

This material is based upon High Performance Computing (HPC) resources supported by the University of Arizona TRIF, UITS, and Research, Innovation, and Impact (RII) and maintained by the UArizona Research Technologies department.

This study has made use of the Transiting Exoplanet Survey Satellite (TESS).
Funding for the TESS mission is provided by the NASA's Science Mission Directorate.
{All the TESS data used in this paper can be found in MAST: \dataset[10.17909/t9-nmc8-f686]{http://dx.doi.org/10.17909/t9-nmc8-f686} \citep{https://doi.org/10.17909/t9-nmc8-f686}.
}

This work has made use of data from the European Space Agency (ESA) mission \textit{Gaia} (\url{https://www.cosmos.esa.int/gaia}), processed by the \textit{Gaia} Data Processing and Analysis Consortium (DPAC, \url{https://www.cosmos.esa.int/web/gaia/dpac/consortium}). 
Funding for DPAC has been provided by national institutions, particularly those participating in the \textit{Gaia} Multilateral Agreement.

\end{acknowledgments}

\clearpage

\begin{table}[ht]
\centering
\caption{15 stars as the sample sources}
\begin{tabular}{ccc}
\hline
      TIC &  Number of Sectors &  Number of flares \\
\hline
364588501 & 34 & 559 \\
201883033 & 4 & 110 \\
140045538 & 3 & 229 \\
441398770 & 3 & 169 \\
229147927 & 5 & 58 \\
63781635 & 2 & 17 \\
25118964 & 33 & 355 \\
231017428 & 4 & 97 \\
234506911 & 6 & 98 \\
220433363 & 22 & 845 \\
231914259 & 6 & 157 \\
365006789 & 3 & 75 \\
150359500 & 32 & 857 \\
388857263 & 4 & 117 \\  
\hline
15 stars & 183 & 4589\\
\hline
\end{tabular}
\label{tab:16_stars_sample_sources}
\end{table}

\begin{deluxetable}{c c c c c c c c c c c c c c c}
\tabletypesize{\normalsize}
\tablecaption{\\ The estimated values of the classification characteristics of the example profiles shown in Figure~\ref{fig:true_and_false_flare_exp}.}

\tablecolumns{19}
\tablewidth{0.1pt}
\tablehead{
\colhead{} & \colhead{$\delta$t}  &\colhead{}$\delta t_{1} / \delta t_{2}$ &\colhead{$ED_{1} / ED_{2}$} & \colhead{$A_{f}/\sigma(F_{\text{LC}_{1}})$}  & \colhead{Flare?}
\\
\colhead{} & \colhead{days} & \colhead{} & \colhead{} & \colhead{} &\colhead{} & \colhead{}
}
\startdata
1 &    0.074 &  0.082 &  0.146 &           208.987 &  True\\
2 &    0.106 &  0.169 &  0.611 &           135.727 &  True\\
3 &    0.013 &  0.125 &  0.287 &            19.524 &  True\\
4 &    0.008 &  5.000 &  0.435 &             2.963 &  False\\
5 &    0.311 &  2.672 &  0.149 &             4.384 &  False\\
6 &    0.012 &  0.500 &  0.203 &             3.359 &  False\\
\enddata

\tablenotetext{}{{Note:} {The columns are: (1) the duration of the flare ($\delta t$) in days, (2) the impulsive-decay time ratio ($\delta t_{1} / \delta t_{2}$), (3) upper-to-lower ED ratio ($ED_{1} / ED_{2}$), and (4) the peak amplitude signal-to-sigma ratio ($A_{f}/\sigma(F_{\text{LC}_{1}})$). }
}
\label{tab:true_and_false_flare_exp}
\end{deluxetable}

\begin{table}[ht]
\centering
\caption{The best hyperparameters for our DNN training}
\begin{tabular}{llll}
\hline
Parameter                & Best values  \\         
\hline
Number of hidden layers             &  3 \\                     
Number of nodes per layer    &   480 (1st), 192 (2nd), and 512 (3rd) \\             
Learning rate                & 0.001  \\
Batch size                   & 32             \\            
\hline
\end{tabular}
\label{tab:best_hyper_DNN}
\end{table}



\begin{table}[ht]
\caption{The best hyperparameters of our Random Forest model}
\begin{tabular}{lll}
\hline
Parameter       & Description & Best setting  \\         
\hline
\texttt{bootstrap}           & Indicates whether bootstrap samples are used when building trees. & True$^{a}$ \\                     
\texttt{criterion}   & Specifies the function to measure the quality of a split. & entropy$^{b}$ \\  
\texttt{oob\_score} & Using out-of-the-bag samples to compute accuracy     & True$^{c}$ \\
\texttt{n\_estimators}                   & The number of trees in the forest. & 200             \\     
\texttt{max\_depth} & The maximum depth of each tree in the forest. & 34\\
\texttt{max\_leaf\_nodes} & The maximum number of leaf nodes a tree can have. & 500\\
\texttt{min\_impurity\_decrease} & The minimum impurity decrement value to split a node. & 1e-08\\
\texttt{min\_samples\_split} & The minimum number of samples required to split an internal node. & 5 \\
\hline
\end{tabular}     
\tablecomments{}{
(a) If True, each tree in the forest is built on a bootstrap sample from the data.\\
(b) We used the entropy function as described in Eq.~\ref{ep:entropy}, and the total numbers of classes $n = 2$ in our study.\\
(c) If True, out-of-bag error is used to evaluate the model's performance in addition to the cross-validation testing set.}

\label{tab:best_hyper_RF}
\end{table}


\begin{table}[ht]
\centering
\caption{The feature importance distribution of our Random Forest and XGBoost models}
\begin{tabular}{ccccc}
\hline
Feature & \multicolumn{2}{c}{Importance weight} \\
                       & (Random Forest)   & (XGBoost)\\         
\hline
$A_{f}/\sigma(F_{\text{LC}_{1}})$      &  0.58    & 0.66\\                     
$\delta$t    &   0.21   & 0.18\\             
$\delta t_{1} / \delta t_{2}$                & 0.14   & 0.1  \\
$ED_{1} / ED_{2}$                & 0.07       & 0.06      \\            
\hline
\end{tabular}
\label{tab:RF_feature_importance}
\end{table}

\begin{table}[h!]
\caption{The best hyperparameters of our XGBoost Classifier model}
\begin{tabular}{lll}
\hline
\hline
Parameter          & Description                                                                 & Best Setting \\
\hline
\texttt{subsample}          & Subsample ratio of the training instances.                                           & 0.51              \\
\texttt{scale\_pos\_weight} & Balancing of positive and negative weights.                                          & 1                    \\
\texttt{reg\_lambda}        & L2 regularization term on weights (Ridge regression).                                & 2.47               \\
\texttt{reg\_alpha}         & L1 regularization term on weights (Lasso regression).                                & 0.15              \\
\texttt{n\_estimators}      & Number of gradient boosted trees.                                                    & 118                  \\
\texttt{min\_child\_weight} & Minimum sum of instance weight needed in a child.                                    & 3                    \\
\texttt{max\_depth}         & Maximum depth of a tree.                                                             & 10                    \\
\texttt{learning\_rate}     & Step size shrinkage used in update to prevent overfitting.                           & 0.1              \\
\texttt{gamma}              & Minimum loss reduction required to split a leaf node further.          & 0.53           \\
\texttt{colsample\_bytree}  & Subsample ratio of columns when constructing each tree.                              & 0.85               \\
\hline
\end{tabular}
\tablecomments{For \texttt{min\_child\_weight}, the condition is $\sum h_i \geq \text{min\_child\_weight}$, where $h_i$ represents the hessian (second derivative of the loss function with respect to the prediction). The parameter \texttt{gamma} sets the minimum loss reduction needed to split a leaf node further in the tree. In other words, it plays a similar role as the parameter \texttt{min\_impurity\_decrease} in Random Forest.}
\label{tab:xgb_hyperparameters}
\end{table}

\begin{deluxetable}{c c c c c c c c c c c c c c c}
\tabletypesize{\normalsize}
\tablecaption{\\ Performance Four Metrics of Machine Learning Models for Flare Detection. }

\tablecolumns{19}
\tablewidth{0.1pt}
\tablehead{
\colhead{Models} & \colhead{\( Accuracy \)}  &\colhead{\( Precision \)} &\colhead{\( Recall \)} & \colhead{\( F_{1} \) scroe} 
\\
}
\startdata
DNN &  95\% &  94\% &       96\% &  95\% \\
RF &  97\% &  96\% &       97\% &  97\% \\
XGBoost &  98\% &  98\% &       97\% &  98\% \\
Multi-algorithm approach  &  98\% &  98\% &   98\% &  98\%
\enddata

\tablenotetext{}{}
\label{tab:four_metrics_scores}
\end{deluxetable}

\newpage
\begin{deluxetable}{c c c c}
\tabletypesize{\normalsize}
\tablecaption{The coefficients of the formulae, Eq.~\ref{eq:duration_calibration} and Eq.~\ref{eq:energy_amplitude_calibrate}, for calibrating flare duration, amplitude, and energy (see Figure~\ref{fig:flare_parameters_calibration}).} \label{tab:formulae_coefficients_for_parameters_calibration}
\tablecolumns{4}
\tablewidth{0.1pt}
\tablehead{
\colhead{Flare parameter (SNR of light curve)} & \colhead{\( a \)} & \colhead{\( b \)} & \colhead{\( c \)}  
} 
\startdata
Duration (SNR=10) & 1.14$\pm$0.02 & 2.72$\pm$0.23 & ... \\
Duration (SNR=20) & 1.19$\pm$0.01 & 1.61$\pm$0.09 & ... \\
Duration (SNR=100) & 1.14$\pm$0.01 & 1.13$\pm$0.07 & ... \\
Duration (SNR=200) & 1.15$\pm$0.02 & 0.89$\pm$0.07 & ... \\
Duration (SNR=1000) & 1.17$\pm$0.02 & 0.63$\pm$0.04 & ... \\
Duration (SNR=2000) & 1.13$\pm$0.02 & 0.65$\pm$0.05 & ... \\
Amplitude (SNR=10) & 0.39$\pm$0.17 & -0.25$\pm$0.02 & 1.14$\pm$0.01 \\
Amplitude (SNR=20) & 0.39$\pm$0.17 & -0.30$\pm$0.02 & 1.06$\pm$0.00 \\
Amplitude (SNR=100) & 0.41$\pm$0.21 & -0.28$\pm$0.02 & 1.02$\pm$0.00 \\
Amplitude (SNR=200) & 0.29$\pm$0.27 & -0.27$\pm$0.02 & 1.01$\pm$0.00 \\
Amplitude (SNR=1000) & 0.20$\pm$0.34 & -0.26$\pm$0.02 & 1.01$\pm$0.00 \\
Amplitude (SNR=2000) & 0.04$\pm$0.38 & -0.24$\pm$0.03 & 1.01$\pm$0.00 \\
Energy (SNR=10) & -1.11$\pm$0.53 & 0.53$\pm$0.05 & 0.79$\pm$0.00 \\
Energy (SNR=20) & 8.13$\pm$0.94 & 0.32$\pm$0.07 & 0.89$\pm$0.00 \\
Energy (SNR=100) & 57.90$\pm$2.68 & 0.61$\pm$0.18 & 0.98$\pm$0.00 \\
Energy (SNR=200) & 114.91$\pm$6.36 & 0.58$\pm$0.16 & 0.99$\pm$0.00 \\
Energy (SNR=1000) & 1242.92$\pm$206.72 & -224.49$\pm$24.53 & 0.99$\pm$0.00 \\
Energy (SNR=2000) & -25.58$\pm$10.56 & -40.66$\pm$11.35 & 0.99$\pm$0.00 \\
\enddata
\end{deluxetable}

\begin{deluxetable}{c c c c c c c c c c c c c c c}
\tabletypesize{\normalsize}
\tablecaption{\\ Catalog of Flares detected in this study.}\label{tab:flares_catalog}
\tablecolumns{12}
\tablewidth{0.1pt}
\tablehead{
\colhead{TIC} & \colhead{$A_{f}$} & \colhead{Cal. $A_{f}$} & \colhead{$\delta$t} & \colhead{Cal. $\delta$t} & \colhead{$E_{f}$} & \colhead{Cal. $E_{f}$} & \colhead{$ED_{1}/ED_{2}$} & \colhead{$\delta t_{1}/\delta t_{2}$} & \colhead{t$_{start}$} & \colhead{t$_{peak}$} & \colhead{t$_{end}$}\\
\colhead{} & \colhead{} & \colhead{} & \colhead{(min)} & \colhead{(min)} & \colhead{(erg)} & \colhead{(erg)} & \colhead{} & \colhead{} & \colhead{(BTJD)} & \colhead{(BTJD)} & \colhead{(BTJD)}}
\startdata
33905 & 0.05500 & 0.09065 & 10.00 & 15.51 & 2.87e+32 & 2.89e+32 & 0.41 & 0.67 & 1622.735 & 1622.737 & 1622.741 \\
33905 & 0.02626 & 0.05543 & 8.00 & 9.86 & 1.15e+32 & 1.16e+32 & 0.47 & 0.33 & 2357.692 & 2357.694 & 2357.698 \\
33905 & 0.14680 & 0.17275 & 24.00 & 35.05 & 8.52e+32 & 8.57e+32 & 0.19 & 0.20 & 2341.420 & 2341.423 & 2341.437 \\
33905 & 0.03928 & 0.05854 & 12.00 & 15.75 & 2.26e+32 & 2.27e+32 & 0.38 & 0.50 & 2350.532 & 2350.534 & 2350.540 \\
33905 & 0.35359 & 0.48844 & 14.00 & 18.81 & 1.32e+33 & 1.33e+33 & 0.22 & 0.17 & 2357.683 & 2357.684 & 2357.692 \\
33905 & 0.02675 & 0.04499 & 10.00 & 12.76 & 1.50e+32 & 1.51e+32 & 0.43 & 0.25 & 2344.166 & 2344.168 & 2344.173 \\
33905 & 0.02832 & 0.04764 & 10.00 & 12.76 & 1.36e+32 & 1.36e+32 & 0.34 & 0.67 & 2346.014 & 2346.016 & 2346.020 \\
33905 & 0.02391 & 0.05047 & 8.00 & 9.86 & 1.00e+32 & 1.01e+32 & 0.60 & 0.33 & 2347.965 & 2347.966 & 2347.970 \\
33905 & 0.02378 & 0.05020 & 8.00 & 9.86 & 9.47e+31 & 9.52e+31 & 0.52 & 0.33 & 2353.184 & 2353.186 & 2353.190 \\
33905 & 0.02056 & 0.04340 & 8.00 & 9.86 & 9.03e+31 & 9.07e+31 & 0.41 & 0.33 & 2350.877 & 2350.879 & 2350.883 \\
34900 & 0.03264 & 0.03550 & 48.00 & 52.76 & 7.78e+33 & 7.79e+33 & 0.27 & 0.33 & 2339.687 & 2339.695 & 2339.720 \\
34900 & 0.00500 & 0.00710 & 14.00 & 13.04 & 5.57e+32 & 5.58e+32 & 0.23 & 0.75 & 2348.025 & 2348.029 & 2348.034 \\
34900 & 0.00393 & 0.00482 & 22.00 & 21.77 & 8.71e+32 & 8.73e+32 & 0.55 & 0.22 & 2336.268 & 2336.270 & 2336.283 \\
34900 & 0.00737 & 0.00828 & 36.00 & 38.07 & 2.39e+33 & 2.40e+33 & 0.36 & 0.38 & 2355.438 & 2355.445 & 2355.463 \\
34900 & 0.00270 & 0.00465 & 10.00 & 8.90 & 2.93e+32 & 2.94e+32 & 0.53 & 0.25 & 2350.326 & 2350.327 & 2350.333 \\
34900 & 0.00451 & 0.00953 & 8.00 & 6.91 & 2.90e+32 & 2.91e+32 & 0.34 & 0.33 & 2358.267 & 2358.269 & 2358.273 \\
34900 & 0.00678 & 0.00878 & 18.00 & 17.34 & 9.96e+32 & 9.98e+32 & 0.31 & 0.50 & 2353.426 & 2353.430 & 2353.438 \\
34900 & 0.00527 & 0.00807 & 12.00 & 10.95 & 6.34e+32 & 6.36e+32 & 0.60 & 0.20 & 2338.532 & 2338.533 & 2338.540 \\
34900 & 0.03725 & 0.04255 & 32.00 & 33.31 & 6.24e+33 & 6.25e+33 & 0.33 & 0.14 & 2352.615 & 2352.618 & 2352.637 \\
40014 & 0.07382 & 0.12167 & 10.00 & 15.51 & 5.04e+31 & 5.08e+31 & 0.35 & 0.25 & 3070.424 & 3070.426 & 3070.431 \\
40014 & 0.04411 & 0.06447 & 12.00 & 19.09 & 3.88e+31 & 3.91e+31 & 0.40 & 0.50 & 2342.947 & 2342.950 & 2342.955 \\
40014 & 0.20600 & 0.22520 & 36.00 & 66.67 & 4.41e+32 & 4.46e+32 & 0.39 & 0.38 & 2335.554 & 2335.561 & 2335.579 \\
40014 & 0.08418 & 0.10857 & 16.00 & 26.49 & 8.09e+31 & 8.16e+31 & 0.26 & 0.33 & 2348.075 & 2348.077 & 2348.086 \\
40014 & 0.09040 & 0.11660 & 16.00 & 26.49 & 8.10e+31 & 8.18e+31 & 0.16 & 0.60 & 3091.841 & 3091.845 & 3091.852 \\
51431 & 0.05158 & 0.06511 & 18.00 & 25.15 & 2.12e+33 & 2.13e+33 & 0.38 & 0.80 & 3089.260 & 3089.266 & 3089.273 \\
51431 & 0.03428 & 0.04495 & 16.00 & 21.95 & 1.31e+33 & 1.31e+33 & 0.48 & 0.33 & 2338.527 & 2338.530 & 2338.539 \\
51431 & 0.01431 & 0.02408 & 10.00 & 12.76 & 4.44e+32 & 4.46e+32 & 0.53 & 0.67 & 2345.708 & 2345.711 & 2345.715 \\
51431 & 0.02335 & 0.03062 & 16.00 & 21.95 & 8.04e+32 & 8.08e+32 & 0.19 & 0.33 & 2337.744 & 2337.747 & 2337.755 \\
\enddata
\tablenotetext{}{TIC: TESS Input Catalog identifier; $A_{f}$: Observed flare amplitude; Cal. $A_{f}$: Calibrated flare amplitude; $\delta$t: Observed flare duration in minutes; Cal. $\delta$t: Calibrated flare duration in minutes; $E_{f}$: Observed flare energy in ergs; Cal. $E_{f}$: Calibrated flare energy in ergs; $ED_{1}/ED_{2}$: Ratio of equivalent durations; $\delta t_{1}/\delta t_{2}$: Ratio of flare implsive and decay durations; t$_{start}$: Flare start time in Barycentric TESS Julian Date (BTJD); t$_{peak}$: Flare peak time in BTJD; t$_{end}$: Flare end time in BTJD.
The full machine-readable table is available online.}

\end{deluxetable}

\begin{deluxetable}{c c c c c c c c c c c c c}
\tabletypesize{\normalsize}
\tablecaption{\\Catalog of Flaring Stars in this study.} \label{tab:flaring_stars_catalog}
\tablecolumns{12}
\tablewidth{0.1pt}
\tablehead{
\colhead{TIC} & \colhead{RAJ2000} & \colhead{DEJ2000} & \colhead{$BP-RP$} & \colhead{Gmag} & \colhead{Tmag} & \colhead{Teff} & \colhead{Rad} & \colhead{Dist} & \colhead{P$_{rot}$} & \colhead{V$_{amp}$} & \colhead{R$_{cont}$} & \colhead{N$_{f}$}\\
\colhead{} & \colhead{(deg)} & \colhead{(deg)} & \colhead{} & \colhead{} & \colhead{} & \colhead{(K)} & \colhead{(R$_{\odot}$)} & \colhead{(pc)} & \colhead{(days)} & \colhead{} & \colhead{} & \colhead{}
}
\startdata
33905 & 220.007 & -25.243 & 2.57 & 13.06 & 11.83 & 3394.00 & 0.37 & 39.12 & 2.71 & 0.0273 & 7.884 & 11 \\
34900 & 219.946 & -26.698 & 1.34 & 9.62 & 8.95 & ... & ... & 44.75 & 2.41 & 0.0095 & 0.002 & 10 \\
40014 & 220.158 & -26.089 & 3.29 & 14.99 & 13.54 & 3051.00 & 0.20 & 31.93 & 0.40 & 0.0389 & 0.030 & 5 \\
51431 & 220.522 & -26.777 & 1.97 & 12.66 & 11.67 & 3796.00 & 0.62 & 82.94 & 5.91 & 0.0709 & 1.158 & 19 \\
51432 & 220.525 & -26.778 & 1.75 & 12.21 & 11.37 & 4005.00 & 0.69 & 82.43 & 5.89 & 0.0570 & 0.700 & 14 \\
58192 & 220.717 & -24.492 & 0.87 & 7.61 & 7.16 & 5580.50 & 1.93 & 68.92 & ... & 0.0010 & 0.003 & 4 \\
67228 & 221.088 & -25.895 & 0.98 & 8.84 & 8.33 & 5484.00 & 1.46 & 74.63 & 5.18 & 0.0057 & 0.006 & 5 \\
68069 & 221.062 & -27.290 & 3.09 & 14.10 & 12.70 & 3140.00 & 0.24 & 30.39 & 0.38 & 0.0359 & 0.019 & 14 \\
70111 & 221.204 & -29.699 & 2.50 & 12.94 & 11.72 & 3434.00 & 0.48 & 66.46 & 0.52 & 0.0151 & 0.271 & 20 \\
82063 & 221.502 & -25.444 & 0.78 & 6.93 & 6.52 & 5910.00 & 1.03 & 30.70 & ... & 0.0337 & 4.733 & 7 \\
102723 & 222.133 & -26.474 & 2.52 & 12.73 & 11.52 & 3421.00 & 0.42 & 38.75 & 4.05 & 0.0184 & 0.039 & 23 \\
134947 & 223.047 & -28.278 & 2.79 & 13.93 & 12.64 & 3286.00 & 0.34 & 46.66 & 1.35 & 0.0298 & 0.036 & 22 \\
146287 & 223.323 & -24.967 & 2.55 & 13.28 & 12.06 & 3404.00 & 0.39 & 47.45 & ... & 0.0131 & 0.012 & 9 \\
590241 & 71.643 & -26.465 & 3.06 & 14.97 & 13.75 & 3170.00 & ... & 82.04 & 0.88 & 0.0466 & 0.461 & 2 \\
593228 & 71.846 & -27.843 & 2.65 & 10.45 & 9.22 & 3356.00 & 0.47 & 15.60 & 6.95 & 0.0105 & 2.267 & 28 \\
\enddata
\tablenotetext{}{TIC: TESS Input Catalog identifier; RAJ2000: Right Ascension in J2000 coordinates (degrees); DEJ2000: Declination in J2000 coordinates (degrees); $BP-RP$: Gaia BP-RP color index; Gmag: Gaia G-band magnitude; Tmag: TESS magnitude; Teff: Effective temperature (K); Rad: Radius (solar radii); Dist: Distance (in parsecs); P$_{rot}$: Rotation period (in days); V$_{amp}$: light curve variation index; {R$_{cont}$: the contamination ratio as described in Section~\ref{subsec:flare candidates identification}}  ; N$_{f}$: Number of flares detected.
{The uncertainties of P$_{rot}$ and V$_{amp}$ will be included in the machine-readable table.}
The full machine-readable table is available online.}
\end{deluxetable}

\begin{deluxetable}{c c c c c c}
\tabletypesize{\normalsize}
\tablecaption{Power-law indices ($\alpha$) of Flare Frequency Distributions in this study and literature} \label{tab:power-law-FFD}
\tablecolumns{6}
\tablewidth{0pt}
\tablehead{
\colhead{Spectral Type} & \colhead{This Study} & \colhead{Lin2019$^{a}$} & \colhead{Y+L2019$^{b}$} & \colhead{Yang2023$^{c}$} & \colhead{B+E2020$^{d}$}
}
\startdata
A & 1.79$\pm$0.22 & ...           & 1.12$\pm$0.08 & 1.86$\pm$0.15 & 1.26$\pm$0.56  \\
F & 1.74$\pm$0.20 & ...           & 2.11$\pm$0.09 & 2.28$\pm$0.20 & ... \\
G & 1.76$\pm$0.19 & 2.01$\pm$0.01 & 1.96$\pm$0.04 & 1.77$\pm$0.40 & ... \\
K & 1.79$\pm$0.17 & 1.86$\pm$0.02 & 1.78$\pm$0.02 & 1.81$\pm$0.17 & ... \\
M & 1.93$\pm$0.11 & 1.82$\pm$0.02 & 2.13$\pm$0.05 & 1.88$\pm$0.16 & ... \\
\enddata
\tablenotetext{}{$^{a}$~\cite{2019ApJ...873...97L}, $^{b}$~\cite{2019ApJS..241...29Y},$^{c}$~\cite{2023A+A...669A..15Y},$^{d}$~\cite{2020ApJ...905..110B}}
\end{deluxetable}

\begin{deluxetable}{c c c c c c}
\tabletypesize{\normalsize}
\tablecaption{The transition turnoff rotation period and the unsaturated slope in the flare energy-rotation period relationships for F, G, K, and M dwarfs.} \label{tab:flare_vs_rotation}
\tablecolumns{3}
\tablewidth{0pt}
\tablehead{
\colhead{Spectral Type} & \colhead{Transition turnoff P$_{rot}$} & \colhead{Unsaturated slope} & \\
\colhead{ } & \colhead{(days)} & \colhead{ }
}
\startdata
F & 0.75$\pm$0.25 & -0.76$\pm$0.27 \\
G & 2$\pm$0.25 & -0.80$\pm$0.11 \\
K & 3$\pm$0.25 & -1.37$\pm$0.10 \\
M & 8$\pm$0.25 & -1.67$\pm$0.21 \\
\enddata
\end{deluxetable}

\vspace{5mm}






\clearpage
\software{Astropy \citep{2013A&A...558A..33A,2018AJ....156..123A, 2022ApJ...935..167A}, Lightkurve \citep{2018ascl.soft12013L}, Matplotlib \citep{2007CSE.....9...90H},  NumPy \citep{harris2020array}, Pandas \citep{mckinney2010data}, Scipy \citep{2020NatMe..17..261V}, Tensorflow \citep{tensorflow2015-whitepaper}, Keras \citep{chollet2015keras}}
\bibliography{manuscript}
\bibliographystyle{aasjournal}



\end{document}